%% file: ms.tex
\documentclass[iop,revtex4]{emulateapj}

\usepackage{graphicx}
\usepackage{epstopdf}
\usepackage{color}
\usepackage{amsmath}
\allowdisplaybreaks[1]
\begin{document}
\title{Stellar populations in the central 0.5 pc of the Galaxy I: a new method for constructing luminosity functions and surface-density profiles}
\author{T. Do\altaffilmark{1,2,7}, J. R. Lu\altaffilmark{3}, A. M. Ghez\altaffilmark{4}, M. R. Morris\altaffilmark{4}, S. Yelda\altaffilmark{4}, G. D. Martinez\altaffilmark{1}, S. A. Wright\altaffilmark{2,5}, and K. Matthews\altaffilmark{6} }
\altaffiltext{1}{Physics and Astronomy Department, University of California, Irvine, CA 92697}
\altaffiltext{2}{Dunlap Institute for Astronomy and Astrophysics,3.5in
University of Toronto, 50 St. George Street, Toronto M5S 3H4, ON, Canada}
\altaffiltext{3}{Institute for Astronomy, University of Hawaii, HI}
\altaffiltext{4}{Physics and Astronomy Department, University of California,
    Los Angeles, CA 90095-1547}
\altaffiltext{5}{Astronomy Department, University of Toronto, Toronto, ON, Canada}
\altaffiltext{6}{California Institute of Technology, Pasadena, CA}
\altaffiltext{7}{Dunlap Fellow}
\email{do@di.utoronto.ca}

\begin{abstract}
We present new high angular resolution near-infrared spectroscopic
observations of the nuclear star cluster surrounding the Milky Way's central
supermassive black hole. Using the integral-field spectrograph OSIRIS on Keck
II behind the laser-guide-star adaptive optics system, this spectroscopic survey
enables us to separate early-type (young, 4-6 Myr) and late-type (old, $> 1$ Gyr) stars
with a completeness of 50\% down to $K^\prime = 15.5$ mag, which corresponds to $\sim 10$ M$_\odot$ for the early-type stars. This work increases the radial extent of reported OSIRIS/
Keck measurements by more than a factor of 3 from 4$\arcsec$ to 14$\arcsec$ (0.16 pc to
0.56 pc), along the projected disk of young stars. 
For our analysis, we implement a new method of completeness correction using a combination of star-planting simulations and Bayesian inference. We assign probabilities for the spectral type of every source detected in deep imaging down to K$^\prime$ = 15.5 mag using information from spectra, simulations, number counts, and the distribution of stars.
The inferred radial surface-density profiles, $\Sigma(R) \propto R^{-\Gamma}$, for the young stars and late-type giants are consistent with earlier results ( $\Gamma_{early} = 0.93 \pm 0.09$,
$\Gamma_{late} = 0.16 \pm 0.07$). The late-type surface-density profile is approximately flat out to the edge of the survey. While the late-type stellar luminosity function is consistent with the Galactic bulge, the completeness-corrected luminosity function of the early-type stars has significantly more young stars at faint magnitudes compared to previous surveys
with similar depth. This luminosity function indicates that the corresponding
mass function of the young stars is likely less top-heavy than that inferred from previous surveys. 

\end{abstract}


\section{Introduction}

The star cluster at the center of the Milky Way has been observed extensively in the past owing to its unique position in the closest galactic nucleus. The study of its properties has led us to unique insights about its stellar population and has demonstrated the existence of a supermassive black hole at the Galactic center \citep[e.g.][]{2006ApJ...643.1011P,2008ApJ...689.1044G,2009ApJ...692.1075G}. The nuclear star cluster is composed mainly of a massive old stellar cluster with a half-light radius of 5-10 pc \citep{2011ASPC..439..222S}. At the center of the cluster, located within the central $\sim0.5$ pc, is a concentration of young stars (of age 4-6 Myr old) that dominates the luminosity of this region. These two components provide us with different probes of the physical conditions near a supermassive black hole. The presence of the young stars in the strong tidal field of the black hole allows us to study star formation in an extreme environment and provides a test of the universality of the initial mass function. The late-type old stars, on the other hand, provide us with a test of the long-term interactions between a star cluster and a supermassive black hole. This has implications for black hole growth as well as the inward migration of compact objects. 

Our understanding of the nuclear star cluster in the Galactic center is driven in large part by progressively more advanced observing capabilities. Seeing-limited observations of the Galactic center in the infrared enabled the identification of the nuclear star cluster as a peak in the stellar density toward the center of the Galaxy \citep[e.g.][]{1968ApJ...151..145B}. Subsequent spectroscopy led to the discovery that the center of the cluster also hosts a number of bright emission line stars \citep{1991ApJ...382L..19K}, indicating that a population of young stars resides within the central $\sim 0.5$ pc. However, because of the high density of stars in this region, it was not possible to disentangle the two populations of stars through seeing limited observations. Spectroscopy of the spatially integrated light showed a decrease in CO equivalent width toward the center of the cluster, which can be due to either a decrease in the number of red giants or contamination of the spectra by the bright Wolf-Rayet (WR) stars in the region \citep{1996ApJ...456..194H}. 

These limitations were greatly alleviated by the advent of adaptive optics, which allowed diffraction-limited imaging and spectroscopy in the near-infrared on 8-10 m class telescopes. Adaptive optics imaging enabled measurements of the number counts of stars as well as their proper motions in the plane of the sky \citep[e.g.][]{2003ApJ...594..812G,2005ApJ...620..744G}. Integral-field spectroscopy provided the crucial ability to separate the population of young stars from that of the old red giants, thus enabling the study of the two populations independently \citep[e.g.][]{2005ApJ...628..246E,2009ApJ...703.1323D}. In terms of studying the characteristics of star and cluster formation in this region, these advances provide two key observables: the surface-density profile and the luminosity function of the cluster. 

The surface-density profile provides one of the observable features of the dynamical state of the cluster. Early in the formation of the cluster, the stellar distribution reflects its origin; for example, about half of the young stars are observed to be distributed in a thin, clockwise-rotating stellar disk with a steep projected radial surface-density profile of $\sim 1/R^2$ in the disk plane \citep{2006ApJ...643.1011P,2009ApJ...690.1463L}. This may be indicative of their in-situ formation in an accretion disk \citep{2003ApJ...590L..33L}. On the other hand, over time, the cluster will become dynamically relaxed with respect to the black hole and settle into a steady-state density profile, with all traces of its origin removed. \citet{1977ApJ...216..883B} predicted that star clusters with a massive black hole should contain a cusp with a spatial density profile of $r^{-7/4}$ to $r^{-3/2}$, depending on whether the cluster has a single mass population or contains multiple mass components. This property helped to facilitate calculations such as the growth of black holes by stars and the in-spiral rate of compact stellar remnants in galactic nuclei, as the power-law exponent is one the most uncertain parameters for describing the distribution of stars \citep[e.g.][]{2010ApJ...708L..42P}. While the red giants at the Galactic center ($> 1$ Gyr) may have had time to dynamically relax, they unexpectedly show a core-like (i.e., flat) surface-density profile \citep{2009AA...499..483B,2009ApJ...703.1323D}. At present, the origin of the flattening of the surface-density profile is unclear. The flat core could arise from secular evolution of the cluster, such as resonant relaxation, or caused by a drastic event such as the in-fall of another massive black hole \citep{2010ApJ...718..739M,2011ASPC..439..189M,2011ApJ...738...99M}. Increasingly refined measurements of the properties of the structure of the old stellar population, such as its spatial density profile and core radius, are necessary to make progress. 

The luminosity function is one of the most fundamental observable parameters of any stellar population. It is a measure of the relative distribution of stellar luminosities and can be used to determine such properties as the age, star formation history, and initial mass function (IMF) of the cluster.  Much of the early work on the near-infrared luminosity function at the Galactic center was aimed at understanding the old population of bright giants, as many of them can be spatially resolved with seeing-limited imaging and spectroscopy. \citet{1996ApJ...470..864B} conducted one of the the most complete near-IR photometric surveys of the central 2$\arcmin$ ($\sim 5$ pc) of the Galactic center possible under seeing-limited conditions. Their observed luminosity function reached $K \sim 12.5$, where stellar crowding started to dominate. They found that down to these magnitudes, the K luminosity function in this region is consistent with that found by \citet{1995AJ....110.2788T} for Baade's Window, a low extinction region several degrees from the Galactic center. While this suggests that the Galactic center may have the same star formation history and composition as the inner bulge, the observations were not deep enough to reach the red clump at K = 15.5, where most of the red giants at the Galactic center are manifested. 

The IMF is one of the most important observational signatures that connect star formation theories with observations \citep{2007ARAA..45..565M}. As most observations from the local universe show a remarkably consistent IMF across different star formation environments, there is substantial interest in whether the stellar IMF is universal, especially in extreme environments like the Galactic center \citep[see review from][]{2010ARAA..48..339B}. The best population for constraining the IMF of stars at the Galactic center lies in the young stars within the central parsec, which have recently become observationally accessible through integral-field spectroscopy behind AO \citep[e.g.][]{2005ApJ...628..246E,2009ApJ...703.1323D}.  Using the AO-fed IFU SINFONI at VLT, \citet{2006ApJ...643.1011P} were the first to construct a K luminosity function from a spectroscopically selected sample of young stars. Their sample was largely limited to stars brighter than $K = 13$ mag, which is near the transition between evolved OB stars and main-sequence B stars. This corresponds to measuring the mass function only for the evolved massive stars. It is necessary to observe the main sequence for reliable mass function measurements as the stellar atmosphere and evolutionary models may have large uncertainties for the massive evolved stars \citep{2007AA...468..233M}.  \citet{2009ApJ...703.1323D} went deeper with the OSIRIS spectrograph on Keck II to $K < 15.5$, which provided a sample of the early-type main sequence B stars. However, they did not attempt to derive the mass function from the observed luminosity function. \citet{2010ApJ...708..834B} also achieved a similar depth for spectroscopic observations in fields sampling out to $\sim 1$ pc from the center, largely perpendicular to the disk of young stars. They used star planting simulations to derive a completeness correction for the K luminosity function within projected radius $0.\arcsec8 < R < 12\arcsec$ (0.03 to 0.5 pc) and through stellar population synthesis modeling, concluded that the young stars have a very top-heavy IMF, with $dN/dm \propto m^{-0.45\pm0.3}$, compared to a Salpeter IMF of $dN/dm \propto m^{-2.35}$ \citep{2010ApJ...708..834B}. In comparison, the young stars in the central $0.\arcsec8$ (sometimes called the S-stars) show a slope that is consistent with Salpeter, with $\Gamma = -2.15\pm0.3$ for stars with $K > 14.0$ mag. There are suggestions that the S-stars may not originate from the same star formation event that formed the young stars further due to the lack of stars more massive than early B main sequence stars in this region \citep[see review in][]{2010RvMP...82.3121G}. Some theoretical studies however suggest that the S-stars might have originated further out, and were brought in by a combination of dynamical events \citep[e.g.,][]{2009ApJ...697L..44M}. One challenge in interpreting these luminosity functions is that at a depth of $K^\prime < 15.5$, there is only a limited magnitude range from the start of the main sequence at $K^\prime \approx 14.0$; this limited range means that the faintest magnitude bin has a large impact on the slope of the luminosity function, and hence the mass function. 

\begin{figure}[htp]
\center
\includegraphics[angle=90,width=3.5in]{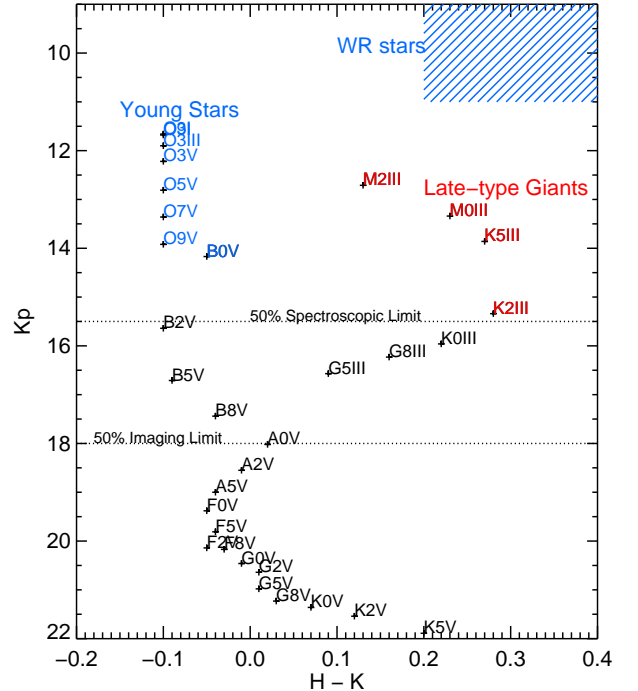}
\caption{Theoretical color-magnitude diagram of the Galactic center in the NIR showing the expected observed K$^\prime$ magnitude for stars of different spectral types behind 2.7 magnitudes of extinction at K$^\prime$ and at 8 kpc (the H-K colors are intrinsic colors). The locations at 50\% completeness for spectroscopy and imaging are also shown. Most of the stars observable with spectroscopy in this region are either young stars (blue) or late-type giants (red). The colors and magnitudes in this plot are derived from \citet{1981MNRAS.196..915W,2001ApJ...558..309D,2006AA...457..637M,2007MNRAS.374.1549W}.}
\label{fig:nir_cmd}
\end{figure}

Because of the importance of the faint end of the luminosity function, it is crucial to understand and carefully account for the assumptions inherent in completeness correction. Completeness correction attempts to characterize an underlying population, in which only a certain number of sources can be observed or identified. In the case of the luminosity function of young stars at the Galactic center, the aim would be to quantify the number of young stars, given that not all sources will have spectral identification. One standard approach is through star-planting simulations as \citet{2010ApJ...708..834B} have done in order to characterize the recovery rate of stars at a given magnitude; the completeness-corrected count is then derived by dividing the observed number of stars by the fraction of undetected sources. This method utilizes no other information than that provided by the star counting simulation. However, for the case of the Galactic center, we have much more information on the underlying population that can be used in the completeness correction: (1) the sources in the magnitude range where spectroscopy is possible have been extensively imaged and their counts are nearly 100\% complete; this means that at a minimum, the completeness-corrected number of sources should not exceed the number of existing sources. (2) Given the luminosity function of stars at the Galactic center at the current spectroscopic sensitivity, we are mainly sensitive to two types of sources - old late-type giants and young stars (Figure \ref{fig:nir_cmd}). (3) The surface density profiles of the early and late-type stars also give information on their expected fraction as a function of projected distance from Sgr A*.

Here, we present new integral-field spectroscopic observations that extend to a distance of $\sim0.5$ pc from Sgr A* in the direction along the projected major axis of the plane of the young stellar disk, with a sky position angle (PA) of 105 deg \citep[e.g.][]{2006ApJ...643.1011P,2009ApJ...690.1463L}. Previously, spectroscopic coverage  beyond about 0.25 pc in this region was limited to lower spatial resolution measurements \citep[e.g.][]{2006ApJ...643.1011P}. We 
obtain spectra for about 400 stars with $K^\prime < 15.5$ mag, which allow us to investigate the radial profile and luminosity functions of both the early-type (young) and late-type (old) populations. In Sections \ref{sec:obs} and \ref{sec:data_reduction}, we describe the new observations and data reduction, while in Section \ref{sec:measurement} we review our method for assigning spectral types. In Section \ref{sec:bayesian}, we adapt a method from Bayesian inference with star-planting simulations to infer probabilities for the spectral types of all sources with K$^\prime < 15.5$ and to establish the spectral completeness of our survey. This incorporates all available information from spectra, number counts, and knowledge of the radial distribution of stars at the Galactic center to estimate the stellar population.  In Section \ref{sec:results}, we present the resulting surface number density profiles and K$^\prime$ luminosity functions of the early and late-type stars and in Section \ref{sec:discussion}, we discuss the implications of these results for the mass function of the young stars and their origins, and we present the implications for the radial distribution of old stars. In a companion paper \citep[Paper II,][]{lu2012} we will derive the mass function of the young stars. Section \ref{sec:summary} summarizes our conclusions.

\section{Observations}
\label{sec:obs}

Observations of the central 0.5 parsec of the Galaxy consist of (1) spectroscopy to distinguish young, hot stars from old, cool giants and (2) photometry to measure the brightness of each young star. Both the spectroscopic and imaging observations were obtained in conjunction with the laser-guide-star adaptive optics (LGS AO) system on the Keck II telescope \citep{2006PASP..118..297W,2006PASP..118..310V}; the laser guide star was propagated at the center of the field of view for each observation, and for low-order tip-tilt corrections, we used the R=13.7 mag star, USNO 0600-28577051, which is located $\sim$19$\tt''$ from SgrA*.   Details specific to the spectroscopic and imaging observations are described in Sections \ref{sec:obs_spec} and \ref{sec:obs_phot}, respectively.

\subsection{Spectroscopy}
\label{sec:obs_spec}
\begin{figure*}[htp]
\center
\includegraphics[angle=90,width=6.5in]{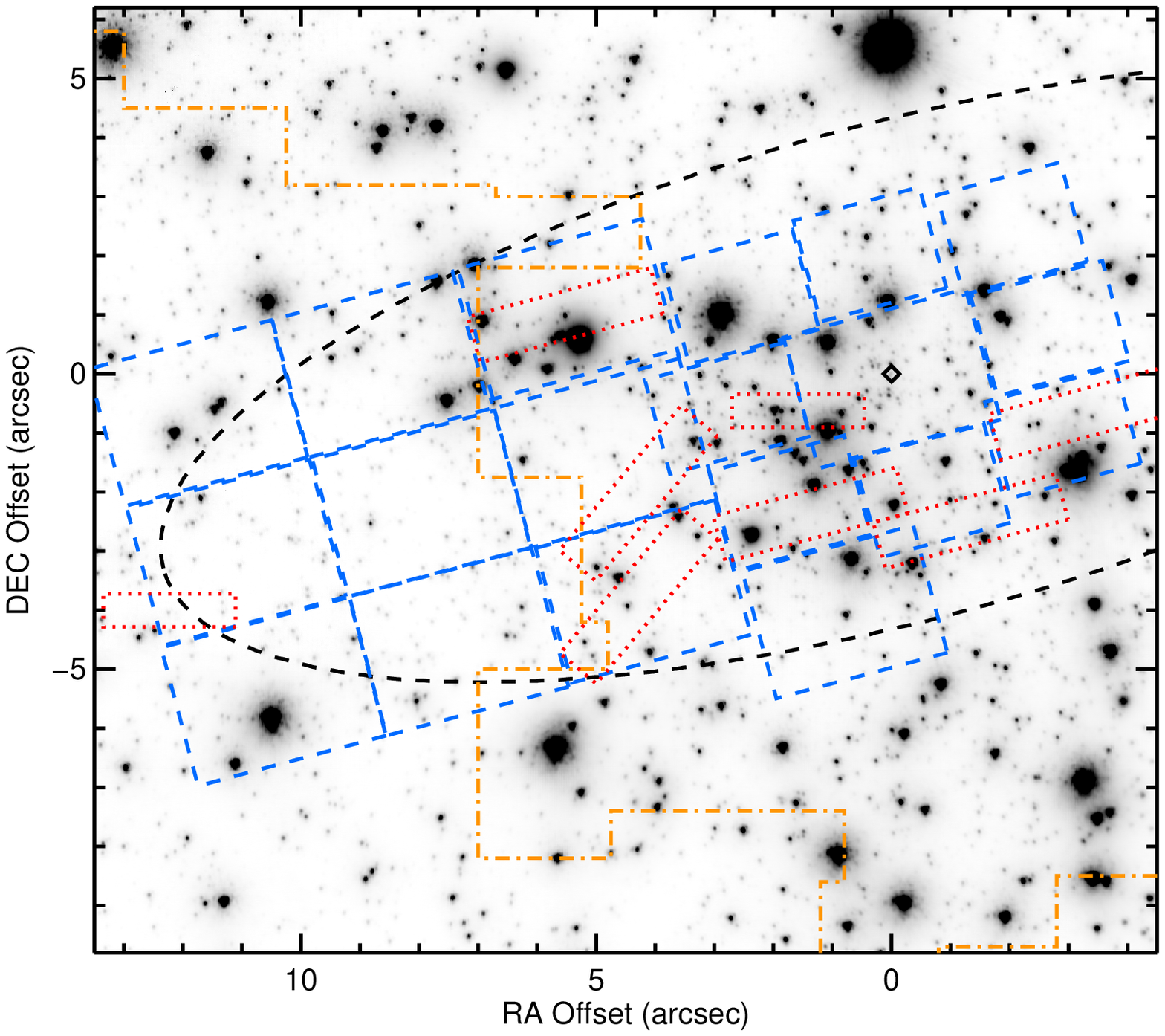}
\caption{The spatial coverage of the OSIRIS survey fields in the Kn3 filter (dashed blue) as well as the K broad-band spectral verification fields (dotted red). The survey is designed to increase the radial coverage along the orientation of the young stellar disk on the plane of the sky \citep[dashed black, $\Omega = 105$ deg, $i = 115$ deg;][]{2009ApJ...690.1463L}. The survey extends radially out to a projected distance of $\sim 0.5$ pc (dashed black) from Sgr A* (black diamond). The survey region from \citet[][dot-dashed orange]{2010ApJ...708..834B} overlaps the GCOWS fields closest to Sgr A*.}
\label{fig:fields_locations}
\end{figure*}

Near-IR integral-field spectra of the Galactic center were obtained between 2007 and 2011 using the OH-Suppressing Infra-Red Imaging Spectrograph \citep[OSIRIS,][]{2006NewAR..50..362L}. The primary observations for this work constitute a survey through the narrow-band filter Kn3 (2.121 to 2.220 \micron). This includes both observations initially reported in  \citet{2009ApJ...703.1323D} and \citet{2010PhDTtdo} and new 2010-2011 observations, which increase the radial extent of this survey by more than a factor of 3 from 4$\arcsec$ to 14$\arcsec$ (0.16 pc to 0.56 pc, see Figure \ref{fig:fields_locations}).  We refer to the combination of the original survey and this new extension of our survey as the Galactic Center OSIRIS Wide-field Survey (GCOWS).  Our initial work, which covered a $8\arcsec\times6\arcsec$ region centered on Sgr A*, used OSIRIS's 35 mas plate scale (field of view of $1.58\arcsec\times2.24\arcsec$). The new GCOWS observations are located at larger projected distances from the Galactic center than the previous work and were obtained with a 50 mas plate scale (field of view of $2.25\arcsec\times3.2\arcsec$); as the stellar densities are lower in this region, this provides a good compromise between spatial resolution and field of view. The new GCOWS fields cover a region of approximately $10\arcsec \times 7.2 \arcsec$ east of the survey reported in \citet{2010PhDTtdo}, along the major axis of the projected disk plane of the clockwise disk of young stars at a PA of 105 degrees, as measured by \citet{2009ApJ...690.1463L} \citep[see also][]{2003ApJ...590L..33L,2006ApJ...643.1011P,2009ApJ...697.1741B,yeldaThesis}. Each of the new fields is observed with a six-point dither pattern of 900 s per frame, in which the dithers have small ($\sim$0\arcsec.1) offsets from one another. The larger plate scale allows us to reach sensitivity comparable to that of the previous 35 mas plate-scale observations, which have about 9 dithers per field. The total surface area of all the observations \citep[including those from][]{2009ApJ...703.1323D} is 113.7 square arcseconds. We also observe 7 pointings (of various total integration times) within the Kn3 survey region with the K broad-band filter (Kbb, 1.965-2.382 $\micron$) in either the 35 mas or 50 mas plate scale, depending on the stellar density. The broad-band observations are used to verify the spectral-types of a sample of stars (see Appendix \ref{sec:verification}). Table \ref{table:obs} summarizes the details of the complete survey, including: field locations, integration times, dates of observations, and data quality.

For calibration purposes, we observe skies after the Galactic center observations. These observations are used to determine the stability of the wavelength solution with the OH sky lines. Sky subtractions for the Galactic center spectra are done using local sky measurements in each of the science data cubes as described in Section \ref{sec:data_reduction}. To remove atmospheric telluric absorption lines, we also observe an A0V (HD195500 or HD155379) and a G2V (HD193193 or HD150437) star each night.

\begin{deluxetable*}{lccccccccccc}
\tablecolumns{12}
\tablecaption{Summary of OSIRIS observations}
\tablewidth{0pc}
\tabletypesize{\scriptsize}  
\tablehead{\colhead{Field Name} & \colhead{Field Center\tablenotemark{a}} & \colhead{Date} & \colhead{N$_{frames}\times t_{int}$} & \colhead{Plate Scale} &  \colhead{FWHM\tablenotemark{b}} & \colhead{Filter} & \colhead{Published\tablenotemark{c}} & \colhead{PA} \\
\colhead{} & \colhead{(\arcsec)} & \colhead{(UT)} & \colhead{(s)}  & \colhead{(mas)} & \colhead{(mas)} & \colhead{} & \colhead{} & \colhead{(degrees)} }
\startdata
GC Central (C)    & 0, 0            & 2008 May 16 & $11\times900$ & 35 & $84\times85$ & Kn3 & 1 & 285 \\ 
GC East (E)      & 2.88, -0.67     & 2007 July 18 & $10\times900$ & 35 & $85\times70$ & Kn3 & 1 & 285\\ 
GC South (S)   & -0.69, -2.00    & 2007 July 19 & $10\times900$ & 35 & $73\times63$   & Kn3 & 1 & 285\\ 
GC West  (W)     & -2.70, 0.74     & 2007 July 20 & $11\times900$ & 35 & $110\times86$ & Kn3 & 1 & 285\\ 
GC Southeast (SE)  & 1.67, -2.23     & 2008 June 03 & $11\times900$ & 35 & $68\times63$ & Kn3 & 1 & 285\\
GC North  (N)    & 0.33, 2.01      & 2008 June 07 & $7\times900$ & 35 & $102\times85$  & Kn3 & 1 & 285\\
              &      & 2008 June 10 & $5\times900$ & 35 & $75\times70$  & Kn3 & 1  & 285\\
GC Northeast (NE)  & 2.55, 1.27      & 2008 June 10 & $5\times900$  & 35 & $74\times68$ & Kn3 & 1 & 285\\
GC Southwest (SW)  & -2.9, -1.12 & 2009 May 26 & $4\times900$ & 35 & $92\times80$ & Kn3 & 2 & 285\\
GC Northwest (NW)  & -1.99, 2.42 & 2009 July 21 & $6\times900$ & 35 & $71\times64$ & Kn3 & 2 & 285\\
E2-1 & 5.43, 0.99 & 2010 May 6 & $6\times900$ & 50 & $94\times96$ & Kn3 & 3 & 285\\
E2-2 & 4.8, -1.4  & 2010 May 7 & $6\times900$ &  50 & $88\times79$ & Kn3 & 3  & 285\\
E2-3 & 4.16, -3.75 & 2010 July 28 & $6\times900$ & 50 & $104\times86$ & Kn3 & 3  & 285\\
E3-1 & 8.59, 0.15 & 2010 May 9    & $6\times900$ & 50 & $79\times86$ & Kn3 &  3 & 285\\
E3-2 & 7.94, -2.21 & 2010 May 7   & $1\times900$ & 50 & $72\times77$ & Kn3 & 3  & 285\\
    &              & 2010 May 9   & $5\times900$ & 50 & $79\times86$ & Kn3 & 3  & 285\\
E3-3 & 7.31, -4.57 & 2010 July 29 & $6\times900$ & 50 & $95\times86$ & Kn3 & 3  & 285\\
E4-1 & 11.73, -0.68 & 2010 May 10 & $6\times900$ & 50 & $97\times84$ & Kn3 & 3  & 285\\
E4-2 & 11.08, -3.04 & 2010 May 9  & $1\times900$ & 50 & $79\times86$ & Kn3 & 3  & 285\\
     &              & 2010 May 10 & $5\times900$ & 50 & $97\times94$ & Kn3 & 3  & 285\\
E4-3 & 10.44, -5.41 & 2010 July 29 & $1\times900$ & 50 & $95\times86$ & Kn3 & 3  & 285\\
     &              & 2010 July 30 & $5\times900$ & 50 & $106\times90$ & Kn3 & 3  & 285\\
S2-1 & 0.69, -4.16  & 2010 August 1 & $6\times900$ & 50 & $102\times85$ & Kn3 & 3  & 285\\
Verification Field 3 & 4.29, -2.20 & 2011 July 25 & $4\times600$ & 50 & 80 & Kbb & 3  & 140 \\
Verification Field 4 & 4.28, -3.76 & 2011 July 25 & $2\times600$ & 50 & 80 & Kbb & 3  & 140 \\
Verification Field 5 & 1.37, -2.38 & 2011 August 17 & $2\times600$ & 50 & 90 & Kbb & 3 & 105 \\
Verification Field 6 & -1.34, -2.49 & 2011 August 17 & $3\times600$ & 50 & 90 & Kbb & 3 & 105  \\
Verification Field 7 & 5.52, 0.99 & 2011 August 17 & $1\times600$ & 50 & 90 & Kbb & 3  & 105 \\
Verification Field 8 & -3.30, -0.65 & 2011 August 17 & $1\times600$ & 50 & 90 & Kbb & 3  & 105 \\
Verification Field 9 & 1.58, -0.62 & 2011 August 25 & $3\times600$ & 35 & 86 & Kbb & 3  & 90 \\
Verification Field 11 & 12.22, -4.00 & 2011 August 25 & $2\times600$ & 35 & 86 & Kbb & 3  & 90\\
\enddata
\tablenotetext{a}{RA and DEC offset from Sgr A* (RA offset is positive to the east).}
\tablenotetext{b}{Average FWHM of a relatively isolated star for the night, found from a two-dimensional Gaussian fit to the source.}
\tablenotetext{c}{1 - \citet{2009ApJ...703.1323D}, 2 - \citet{2010PhDTtdo}, 3 - this work}
\label{table:obs}
\end{deluxetable*}

\subsection{Imaging}
\label{sec:obs_phot}

Photometric observations for the individual stars in the central
parsec of the Galaxy were conducted using the 
K'-band (K$^\prime$) filter at $\lambda_o = 2.12$ \micron~($\Delta \lambda = 0.35$ \micron)
in order to identify stars, measure their positions and K$^\prime$ brightness, and estimate
the completeness of our star counts.
Observations were taken in 2006, 2008,
and 2010 with the NIRC2 instrument (PI: K. Matthews). The NIRC2 field of view is $\sim$10'' with a pixel scale of 9.95 mas/pixel \citep{2010ApJ...725..331Y}.
To image all of the young stars in this region, a mosaic was
constructed covering 27'' $\times$ 27'' roughly centered on Sgr A*. 
Individual exposures at each pointing had an integration time of
$t_{int}=28$ s (2.8 s $\times$ 10 coadds). 
The mosaic dither pattern always consisted of a 3$\times$3 position box pattern with
8\farcs5 steps between each position and with $\sim4$ exposures used at each position\footnote{The observations were typically done under less than ideal seeing conditions and some individual exposures were rejected due to poor AO correction.}.
For 2 of the 3 epochs, we also completed a 2$\times$2 position box pattern
with 4'' steps, and 2 exposures at each position to provide large overlaps between all the tiles in the mosaic.

\section{Data Reduction}
\label{sec:data_reduction}

\subsection{Spectroscopy}

Data reduction and extraction of new spectra were performed in a manner similar to that of \citet{2009ApJ...703.1323D}. For these data, we used version 2.3 of the OSIRIS pipeline, as provided by the instrument team. This version includes a new wavelength solution for the instrument in 2009, which was subject to changes in temperature at that time. This wavelength solution was verified by comparing the locations of OH sky emission lines. The pipeline also removes electronic cross-talk, corrects cosmic rays, and assembles the data cubes. 

Stellar spectra are extracted from the GCOWS Kn3 observations of all stars brighter than a differential extinction corrected K$^\prime_{\Delta A} < $15.5 mag within our field of view, as identified from deep NIRC2 imaging (Section \ref{sec:reduction_photometry}). To register the OSIRIS data cubes to the LGS AO $K^\prime$ images, the point-spread function (PSF) fitting routine, \textit{StarFinder} \citep{2000AAS..147..335D}, is run on an image produced from collapsing the OSIRIS data cube along the spectral dimension, producing an OSIRIS star list. Positions from \textit{StarFinder} are matched to positions derived from LGS AO $K^\prime$ images in the same region. Because the images have higher spatial resolution and better PSF characterization than the OSIRIS observations, they allow us to identify the locations of stars that may have been missed by \textit{StarFinder} on the OSIRIS cubes. Spectra are then extracted with a circular aperture centered at the location of each star (K$^\prime_{\Delta A} < 15.5$) detected in imaging at each spectral channel. We use an aperture radius between 1 and 2 pixels (50-75 mas), depending on the distance to the nearest source. For sky and background subtraction, we use the median flux values in an annulus with an inner radius of 1-2 pixels and an outer radius of 2-4 pixels. To remove atmospheric telluric lines, we divide the spectra by that of a blackbody-removed spectrum of an A star each night. The A star is featureless in the wavelength-region of interest except for the strong Br $\gamma$ line, which we replace by using the spectrum of a G star calibrator divided by the solar spectrum over the region 2.155-2.175 $\micron$. Within the entire GCOWS data set, we extract a total of 400 spectra, including those reported in \citet{2010PhDTtdo}, which are re-extracted and re-analyzed here.

Stellar spectra are extracted from the Kbb observations for 12 stars (described in more detail in Appendix \ref{sec:verification}). For Kbb data taken in the 50 mas plate scale, stellar spectra are extracted in a similar manner to the extraction described above. For the Kbb data taken in the 35 mas plate scale, an aperture radius of 2 pixels is used, with a sky annulus defined from 2-4 pixels.

\subsection{Photometry}
\label{sec:reduction_photometry}
Each tile of the NIRC2 photometric mosaic is reduced separately. 
This is necessary because the
AO point spread function (PSF) varies with time and position and the correct 
PSF is required for precise photometry. 
Our NIRC2 data reduction pipeline is used to 
subtract dark current and sky emission, flatten the field, remove bad
pixels and cosmic rays, and apply corrections for instrumental
and atmospheric distortion \citep{luThesis,2010ApJ...725..331Y}. 
For each tile in the dither pattern, the individual exposures
at that pointing were combined.
Additionally, three subset-images are created for each tile with 
1/3 of the exposures in order to estimate uncertainties.

Stellar photometry and astrometry are extracted using 
{\em StarFinder} with the same setup described in \citet{2010ApJ...725..331Y}.
The resulting starlists are photometrically calibrated using a sample of stellar magnitudes 
reported in \citet{2010AA...511A..18S}, converted from the Ks filter to the K$^\prime$ filter
as described in Appendix \ref{sec:photo_calib}.
Uncertainties are estimated empirically for each tile by taking the error on the mean flux
and position measurements from the tile's three subset-images.
Sources not detected in the tile's combined image and 3 subset images are 
thrown out as spurious artifacts. 
Starlists for all the tiles are then 
mosaicked together to create a single master starlist for each epoch covering the entire
27''$\times$27''. The photometry for stars in the mosaicked starlist is 
the error-weighted average flux of all the tiles in which a star is present. 
The photometric errors are either the weighted standard deviation of the fluxes
in all the tiles, STD$_{weighted}$($f_t$), or the average flux error, AVG($\sigma_{f_i}$), 
whichever is larger. 

The mosaicked starlists from 2006, 2008, and 2010 are aligned together. 
Sources are dropped that are not detected in at least 2 of the 3 epochs, which throws out
most spurious detections due to PSF artifacts and cosmic rays. 
Some stars may have intrinsic brightness variations, so we adopt, as our final 
photometric measurements, the time-averaged flux and RMS error,
weighted by the flux errors at each epoch. Due to the small number of epochs used to estimate the error, we impose a minimum photometric uncertainty of 0.02 mag.

Photometry for stars at the Galactic center must also be corrected for strong and spatially variable extinction, even at near-infrared wavelengths. 
A detailed extinction map has been created for the region from near-infrared photometry 
of red-clump stars by \citet{2010AA...511A..18S}. 
This extinction map is used to apply differential extinction corrections to individual stars, thereby shifting all the observed stars in our NIRC2 imaging to a common extinction value of A$_{Ks}=2.7$, the mean extinction value for the region. 
Before applying the differential extinction correction, we convert our observed K$^\prime$ magnitude to a Ks magnitude using filter conversions computed from a synthetic atmosphere 
with T$_{eff}=$30,000 K for early-type stars, and T$_{eff}$=4,000 K for late-type stars and 
un-typed stars (Appendix \ref{sec:photo_calib}). 
For the untyped sources, the error in assuming the wrong spectral type is less than
the typical photometric error. 
After correcting for extinction, the Ks photometry is converted back to K$^\prime$ magnitudes
and the differential extinction-corrected K$^\prime_{\Delta A}$ photometry is used throughout the paper.

The high stellar density and the large brightness contrast of stars at the Galactic center cause some stars to be undetectable in the NIRC2 images. The imaging completeness as a function of position and brightness is estimated using star planting simulations described in Appendix \ref{sec:image_completeness}. The average resulting completeness is 94\% at K$^\prime_{\Delta A}=$15.5 and 41\% at K$^\prime_{\Delta A} = 18$ mag in the GCOWS field of view (these values are comparable for observed K$^\prime$).

\section{Spectral typing and Bayesian Inference}
\label{sec:measurement}

For the purposes of this study, we wish to differentiate the Wolf-rayet (WR) and O/B stars (main sequence and evolved) from those of later spectral-types such as the evolved M and K giants.  We will refer to early-type stars as stars with a spectral type of B or earlier (including the WR stars) and late-types as all stars with spectral-types later than B. As these two groups of stars were formed at very different times, we can use their early or late-type status as a proxy for age in measurements of the luminosity function and surface-density profiles of the two populations. In this section, we describe our method of classifying stars under the hypotheses that they are either early-type ($H_E$), or late-type ($H_L$). The goal is to assign each star a probability of being early-type, $P_E$, or late-type, $P_L$, and with the constraint that $P_L + P_E$ = 1. This process is composed of the following steps: 

\begin{enumerate}

\item Manually assign each star as either early-type ($P_E = 1$), late-type ($P_L = 1$), or untyped using spectral classification criteria laid out in Section \ref{sec:manual_spec}. 

\item Use the sample of manually typed stars with K$^\prime$ $> 14.0$ to train the Bayesian algorithm to recognize the properties of early and late-type stars. This is accomplished by constructing the probability distributions of Na I and Br $\gamma$ equivalent widths for both the early-type and late-type stars with K$^\prime$$>$14 (Section \ref{sec:training}).

\item For all untyped sources, assign probabilities based on the Bayesian evidence for the early-type and late-type hypotheses using the above training sample and extensive star planting simulations (Section \ref{sec:bayesian}).
\end{enumerate}
For all stars we have extracted (K$^\prime_{\Delta A}$$<$15.5), this analysis yields the probability that each star is either early-type or late-type.

\subsection{Manual Spectral Types}
\label{sec:manual_spec}

We group the stellar spectra by eye into three groups: 1) late-type, 2) early-type, or 3) untyped. Stars with significant Na I features are classified as late-type (219 stars). The sources with Br $\gamma$ absorption and no Na I features are classified as early-type (44 stars). Bright (K$^\prime \lesssim 13.0$) stars with featureless spectra between 2.121 to 2.220 $\micron$ are also classified as early-type (23 stars); these sources are most likely O V or O/B I stars which can have very weak Br $\gamma$ absorption or emission \citep{1996ApJS..107..281H}.  We also identify 12 WR stars, all of which were previously identified by \citet{2006ApJ...643.1011P}, as early-type. The remaining stars with unclear spectral features are classified as untyped; all stars with SNR $< 5$ are also classified as untyped. The above criteria are slightly different from those used in \citet{2009ApJ...703.1323D}. The revision from \citet{2009ApJ...703.1323D} is based on the detection of a few yellow giants in the survey region, which have smaller Na I equivalent widths than the bulk of the K and M giants \citep[some have been identified by][]{2003ApJ...597..323B,2011ApJ...741..108P}.  In the process of revising these criteria, we obtained Kbb spectra of a small subset of sources to verify that this spectral typing method is robust (see Appendix \ref{sec:verification}).   Figure \ref{fig:fields_id} shows the locations of the 286 stars with manually determined spectral classifications. Detailed properties for these late, non-WR early-type, and WR stars are reported in Table \ref{table:late}, Table \ref{table:early}, and Table \ref{table:wr}, respectively\footnote{As in \citet{2009ApJ...703.1323D} we exclude the star S0-32 from our analysis because it is a known foreground source.}. 

\begin{figure*}
\center
\includegraphics[width=6in]{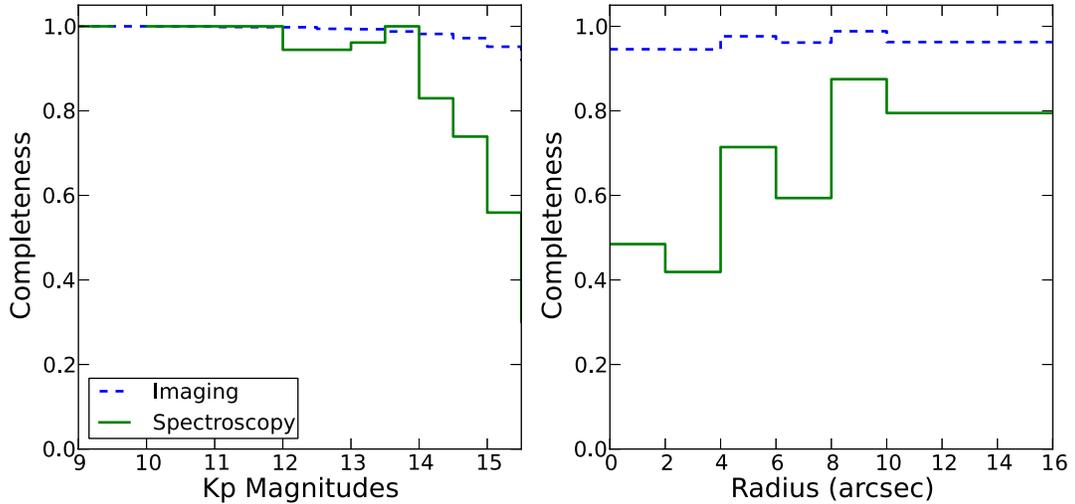}
\caption{
Completeness as a function of brightness ({\em left}) and distance from Sgr A* ({\em right}). The spectroscopic completeness based on manual typing ({\em solid green}) relative to imaging. The imaging completeness based on star planting ({\em dashed blue}) are also shown. Completeness drops below 50\% at K$^\prime_{\Delta A}$=15.5; thus, we only analyze luminosity functions and radial profiles down to this magnitude limit.}
\label{fig:completeness}
\end{figure*}

The sensitivity of the fields observed in 2010 with the 50 mas plate scale is similar to that from \citet{2009ApJ...703.1323D}. Table \ref{tab:field_completeness} summarizes the completeness of each of the new fields compared to imaging in 0.5 mag bins. For the entire GCOWS sample, we are able to spectral-type about 50\% of the sources known from imaging with K$^\prime_{\Delta A}$ between 15.0 to 15.5 mag (Figure \ref{fig:completeness}). The radial dependence of the spectroscopic completeness is shown in Figure \ref{fig:completeness} for $K^\prime = 14.5 - 15.5$ mag. This is compared to the imaging completeness presented in Appendix \ref{sec:image_completeness}. The spectroscopic sensitivity drops dramatically around bright stars (Figure \ref{fig:fields_id}). For example, no stars were spectral-typed at distances closer than 0.\arcsec25 from the IRS sources. Key factors that contribute to the incompleteness of our observations are halo noise from bright stars, background gas emission lines, and crowding in the central regions.

\begin{figure*}[htp]
\center
\includegraphics[angle=90,width=7in]{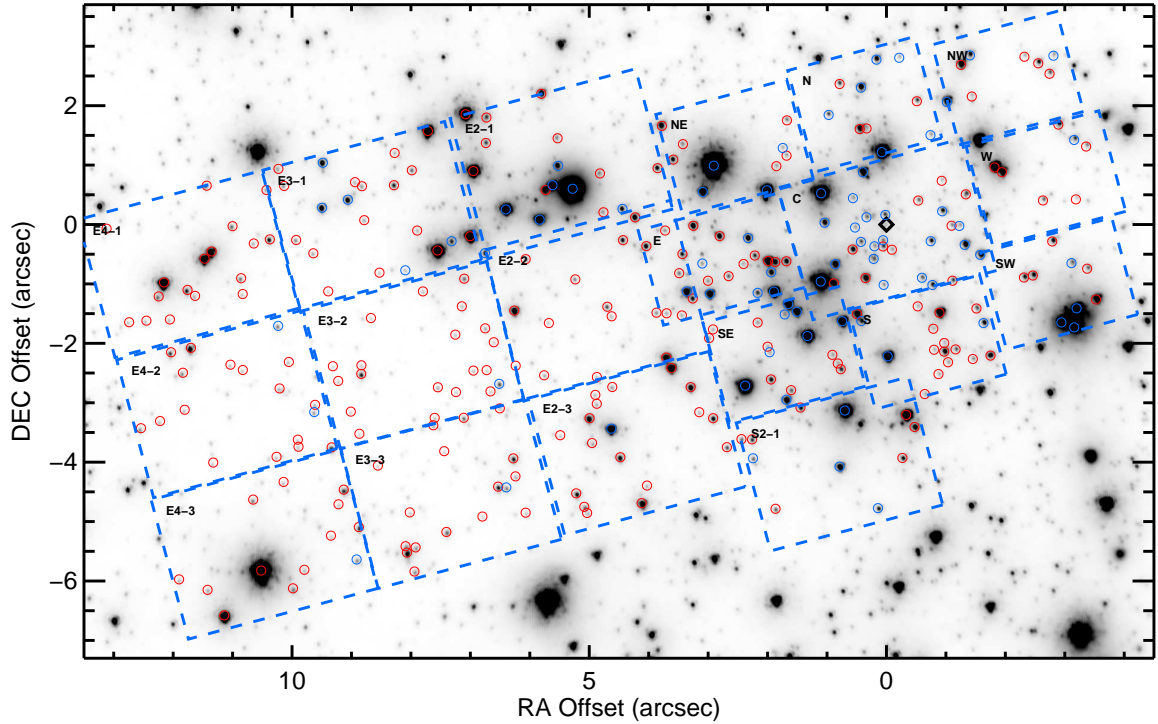}
\caption{Spectral identification of old (red) and young (blue) stars on an LGS AO image of the region. The dashed lines are the outlines of each OSIRIS pointing with field name in each corner. The diamond marks the location of Sgr A*. This is the sample of manually spectral-typed sources.} 
\label{fig:fields_id}
\end{figure*}

\subsection{Training sample: observed line width distributions}
\label{sec:training}
\begin{figure*}[htb]
\center
\includegraphics[angle=90,width=6.5in]{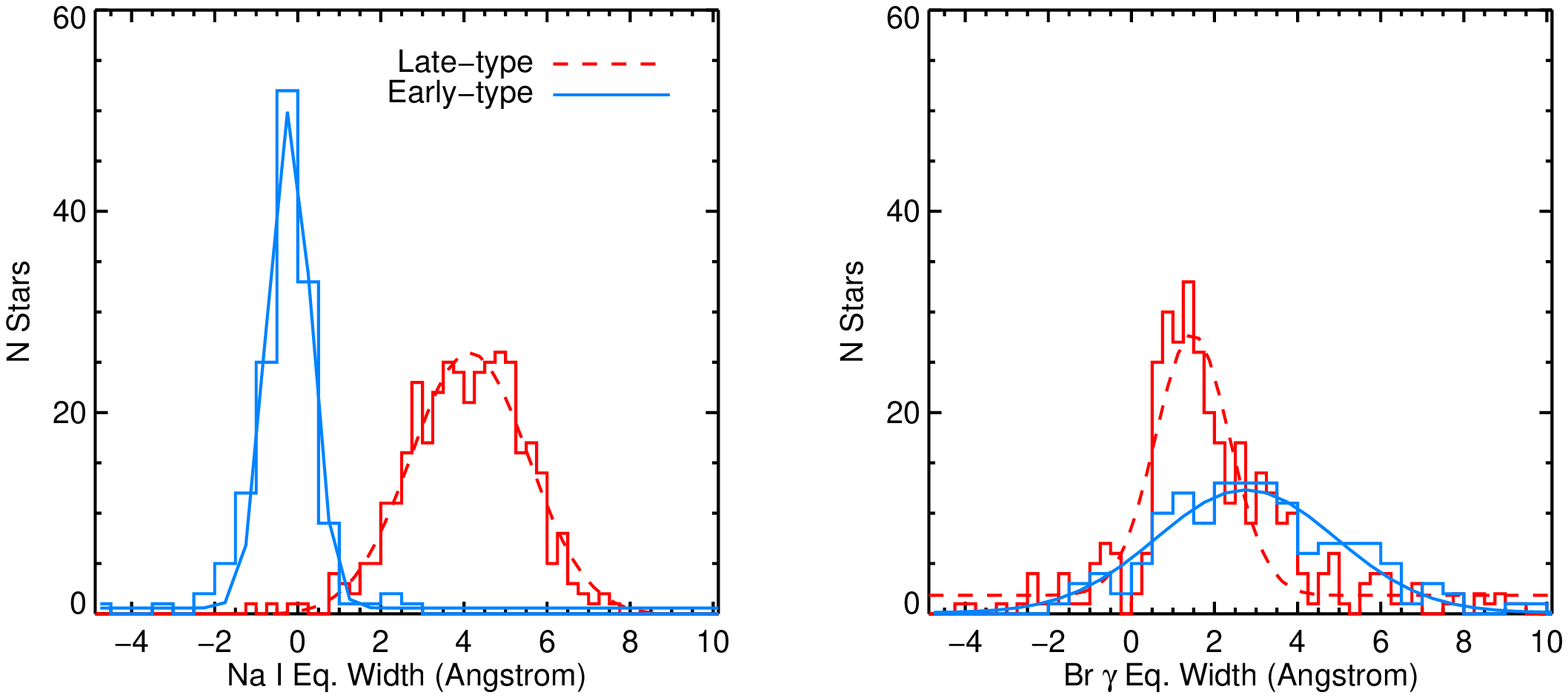}
\caption{ The distributions of equivalent widths for the fainter ($K^\prime > 14$) sample of manually typed early (blue, solid) and late-type (red, dashed) stars, along with the best fit Gaussian distribution for Na I (left) and Br $\gamma$ (right). These distributions are used as priors for the Bayesian inference method of calculating the spectral type probabilities in Section \ref{sec:bayesian}.}
\label{fig:na_br_ew}
\end{figure*}

A key component of assigning probabilities to the untyped stars is understanding the expected distribution of measured Na I and Br $\gamma$ equivalent widths of observable stars at the Galactic center. We use all manually typed stars with K$^\prime$ $>14.0$ to construct a distribution function for the Na I and Br $\gamma$ equivalent widths of early-type and late-type stars (Figure \ref{fig:na_br_ew}); these stars are chosen for the training sample because the majority of the untyped stars have $K^\prime > 14.0$ mag. The method used to measure equivalent widths is described in Appendix \ref{sec:auto_spec_type}. Typical errors in equivalent width of Na I for this sample for the early and late-type stars are about 0.5 and 1 \AA, respectively. The equivalent width errors for Br $\gamma$ area about 0.5 and 0.7 \AA.  A Gaussian was fit to each distribution of equivalent widths (the best fits are summarized in Table \ref{tab:priors}). These Gaussian distributions are used as priors in the following Bayesian analysis of the untyped stars.

\begin{deluxetable}{ccccc}
\tablecolumns{5}
\tablecaption{Observed distribution of spectral line widths for K$^\prime > 14$}
\tablewidth{0pc}
\tabletypesize{\scriptsize}  
\tablehead{\colhead{SpT} & \colhead{Na EW} & \colhead{$\sigma_{Na}$} & \colhead{Br $\gamma$ EW} & \colhead{$\sigma_{Br\gamma}$} \\
\colhead{} & \colhead{(\AA)} & \colhead{(\AA)} & \colhead{(\AA)} & \colhead{(\AA)}}
\startdata
Late-type & 3.7 & 1.3 & 1.9 & 1.2\\
Early-type & -0.2 & 0.6 & 3.3 & 1.8
\enddata
\label{tab:priors}
\end{deluxetable}

\begin{deluxetable*}{ccccccccccc}
\tablecolumns{11}deluxetable*
\tablecaption{GCOWS spectroscopic completeness}
\tablewidth{0pc}
\tabletypesize{\scriptsize}  
\tablehead{\colhead{Mag. Bin\tablenotemark{a}} & \colhead{E2-1\tablenotemark{b}} & \colhead{E2-2} & \colhead{E2-3} & \colhead{E3-1} & \colhead{E3-2} & \colhead{E3-3} & \colhead{E4-1} & \colhead{E4-2} & \colhead{E4-3} & \colhead{S2-1}}
\startdata
\input{gcows_field_completeness}
\enddata
\tablenotetext{a}{The observed K$^\prime$ magnitude bin (not extinction corrected).}
\tablenotetext{b}{The fraction of stars that have been manually spectral-typed out of all sources detected in NIRC2 imaging in the given magnitude bin.}
\label{tab:field_completeness}
\end{deluxetable*}

\subsection{Statistical Spectral Types as an Approach to Completeness Correction}
\label{sec:bayesian}

Spectra of the untyped sources contain important information about the relative completeness of early-type and late-type classifications, which has previously not been incorporated into analyses of luminosity functions and radial density profiles. In the following section, we develop a new approach to completeness correction that utilizes the spectra of untyped sources, along with extensive star planting simulations, to assign each untyped source a probability of being either early-type or late-type. In essence, we compare two hypotheses: (1) the star is late-type ($H_L$) or (2) the star is early-type ($H_E$). The goal is to compare the relative strengths of these two hypotheses and assign a probability for a given star to be early or late-type. To accomplish this goal, the Bayesian evidence is computed for both hypotheses given the observations and our knowledge about the expected spectral features of these sources (Section \ref{sec:bayesian_evidence}). The relative strengths of the hypotheses is the ratio of the Bayesian evidence. To calibrate the Bayesian evidence and determine probabilities, we perform extensive star planting simulations (Section \ref{sec:derive_prob}) and make use of two types of priors (Section \ref{sec:priors}). The results of these analyses are provided in Table \ref{tab:sim_results}.

\subsubsection{Bayesian Evidence}
\label{sec:bayesian_evidence}
First, we consider the Bayesian evidence, which is the likelihood of obtaining observed data, $x$, given a specific hypothesis, $H$, marginalized over all possible model parameters, $\theta$:
\begin{equation}
P(x|H) = \int P(x|\theta,H) P(\theta|H) d\theta 
\end{equation}
$P(x|\theta, H)$ is the likelihood of observing $x$ for a specific model parameter $\theta$, and $P(\theta|H)$ incorporates prior information about the distribution of model parameters. In our case, the Bayesian evidence for the two hypotheses are
\begin{align}
\begin{split}
P(Na_{obs}, Br_{obs} | H_E) = & \int_{-\infty}^{\infty}  P(Na_{obs}, Br_{obs}|Na, Br, H_E) \\
&  \times P(Na, B r| H_E) \; dNa \; dBr
\end{split}\\
\begin{split}
P(Na_{obs}, Br_{obs} | H_L) = & \int_{-\infty}^{\infty}  P(Na_{obs}, Br_{obs}|Na, Br,H_L) \\
& \times P(Na, Br | H_L) \; dNa \; dBr,
\end{split}
\end{align}
where $Na_{obs}$ and $Br_{ons}$ are the Na I and Br $\gamma$ equivalent width measurements for that star. The likelihood functions (e.g. $P(Na_{obs}, Br_{obs}|Na, Br,H_E)$) are assumed to be the product of two independent probability distributions, one for Na I and one for Br $\gamma$,
\begin{equation}
\begin{split}
P(Na_{obs}, Br_{obs} | Na, Br, H_E) = & P(Na_{obs} | Na, H_E) \\
& \times P(Br_{obs} | Br, H_E).
\end{split}
\end{equation}
Each of these terms is modeled as a Gaussian with the observed value as the mean and the error in the observed value as $\sigma_{obs}$. For example, the likelihood for measuring Na EW is:
\begin{equation}
P(Na_{obs}| Na, H_E) = \frac{1}{\sqrt{2\pi}\sigma_{Na_{obs}}}\exp\left[ \frac{-(Na - Na_{obs})^2}{2\sigma_{Na_{obs}}^2}   \right]
\end{equation}
The likelihood functions in this case are not dependent on whether the star is early-type or late-type, since it is only a function of our measurements. The priors on the distributions of $Na$ and $Br$, however, are dependent on the hypothesis. They are also factored into two independent terms, 
\begin{align}
P(Na, Br | H_L) = P(Na | H_L) P(Br | H_L) \\
P(Na, Br | H_E) = P(Na | H_E) P(Br | H_E).
\end{align}
These priors are modeled as Gaussian distributions based on the equivalent width measurements of stars with manually determined spectral types described in Section \ref{sec:manual_spec} (Table \ref{tab:priors}, Figure \ref{fig:na_br_ew}). For example:
\begin{equation}
P(Na|H_E) = \frac{1}{\sqrt{2\pi}\sigma_{Na_{prior}}}\exp\left[ \frac{-(Na- Na_{prior})^2}{2\sigma_{Na_{prior}}^2} \right].
\end{equation}
If the observed value of the equivalent width is far from the peak in the prior, the resulting integral in the evidence will be small, lending less evidence for this hypothesis. 

\begin{figure*}[ht]
\center
\includegraphics[angle=90,width=6in]{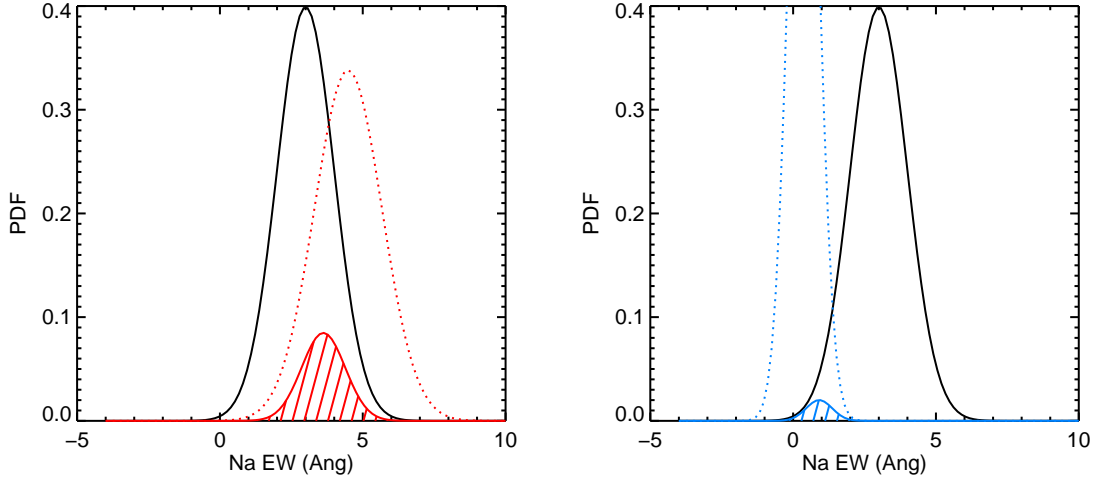}
\caption{An example of calculating the evidence for a late-type star compared to an early-type star. The solid line in the two plots is the measured Na I equivalent width for this star, modeled as a Gaussian with a mean of 3.0 \AA, and  $\sigma = 1.0$ \AA (this is the likelihood for observing the Na I equivalent width). The priors for both the late-type (dotted red line) and early-type (dotted blue line) stars are also shown. The evidence for the hypothesis that the star is late-type or early-type is the integral of the product of the likelihood and the prior. The evidence is the area shown in hashed red and blue, respectively. The Bayes Factor is then the relative area of the two distributions. For this example, as expected, the evidence is quite strong that the star is late-type, with $\ln $(BF) $\approx 2$.}
\label{fig:bayes_factor_example}
\end{figure*}

\begin{figure}[ht]
\center
\includegraphics[angle=90,width=3.5in]{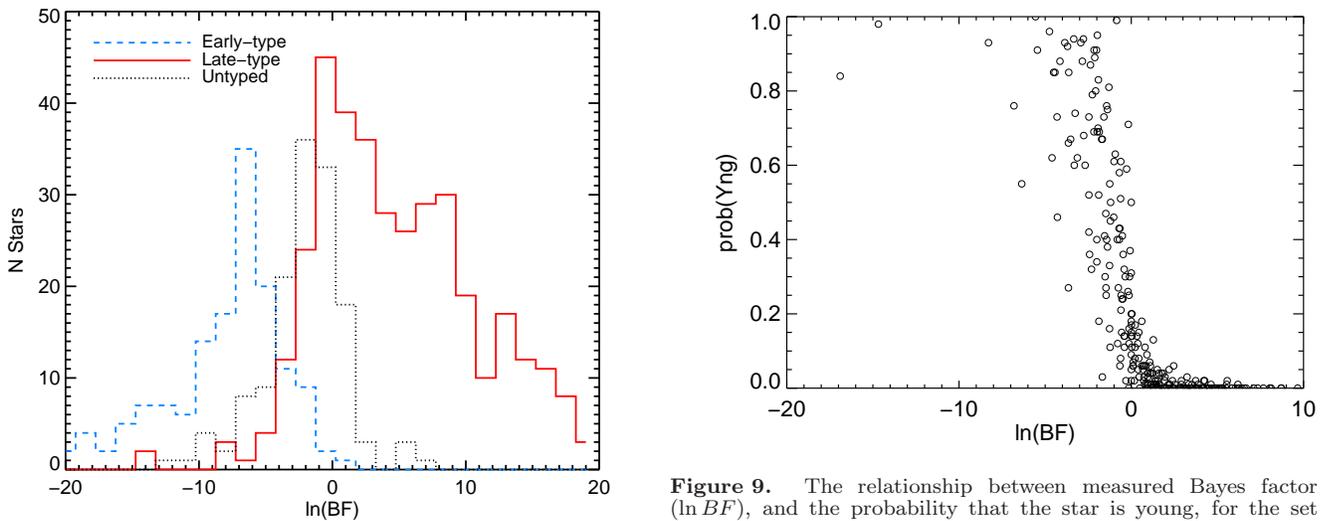}
\caption{The distribution of Bayes factors for stars with manual spectral-types (early-type: dashed blue, late-type: solid red) compared to that of untyped population (dotted black).}
\label{fig:obs_bf_dist}
\end{figure}

The ratio of the evidences for the two hypotheses is the Bayes factor:
\begin{equation}
BF = \frac{P(Na_{obs}, Br_{obs} | H_L)}{P(Na_{obs}, Br_{obs} | H_E) }
\end{equation}
The use of Bayes factors for evaluating the strength of the evidence was first advocated by Jeffereys (1961), and has subsequently been used in cosmology to evaluate different cosmological models \citep[e.g.][]{2008ConPh..49...71T}. A large Bayes factor means that hypothesis $H_L$ is preferred over hypothesis $H_E$, whereas a small value of $BF$ would mean that hypothesis $H_E$ is preferred over $H_L$.  See Figure \ref{fig:bayes_factor_example} for a simplified example of evaluating the relative strength of the evidence (i.e. the Bayes factor) for a late-type star versus an early-type star using only the measured Na I equivalent width. Figure \ref{fig:obs_bf_dist} shows the distribution of $\ln BF$ 

\subsubsection{Deriving Probabilities}
\label{sec:derive_prob}
The observed Bayes Factor for each untyped star is converted into a probability of being early-type ($P_E$) or late-type ($P_L$) by running a series of Monte Carlo simulations. These simulations are used to calibrate the effects of local noise sources, such as mis-subtraction of background Br $\gamma$ gas or halos of nearby bright stars. For each untyped star in our sample, we simulate and plant 100 late-type and 100 early-type stars nearby the source as described in Appendix \ref{sec:sims}. For every planted star, we extract Br $\gamma$ and Na I equivalent widths (Appendix \ref{sec:auto_spec_type}) to compute the Bayes Factors to calibrate how the local environment affect them.

\begin{figure*}[bth]
\center
\includegraphics[width=5.0in,angle=90]{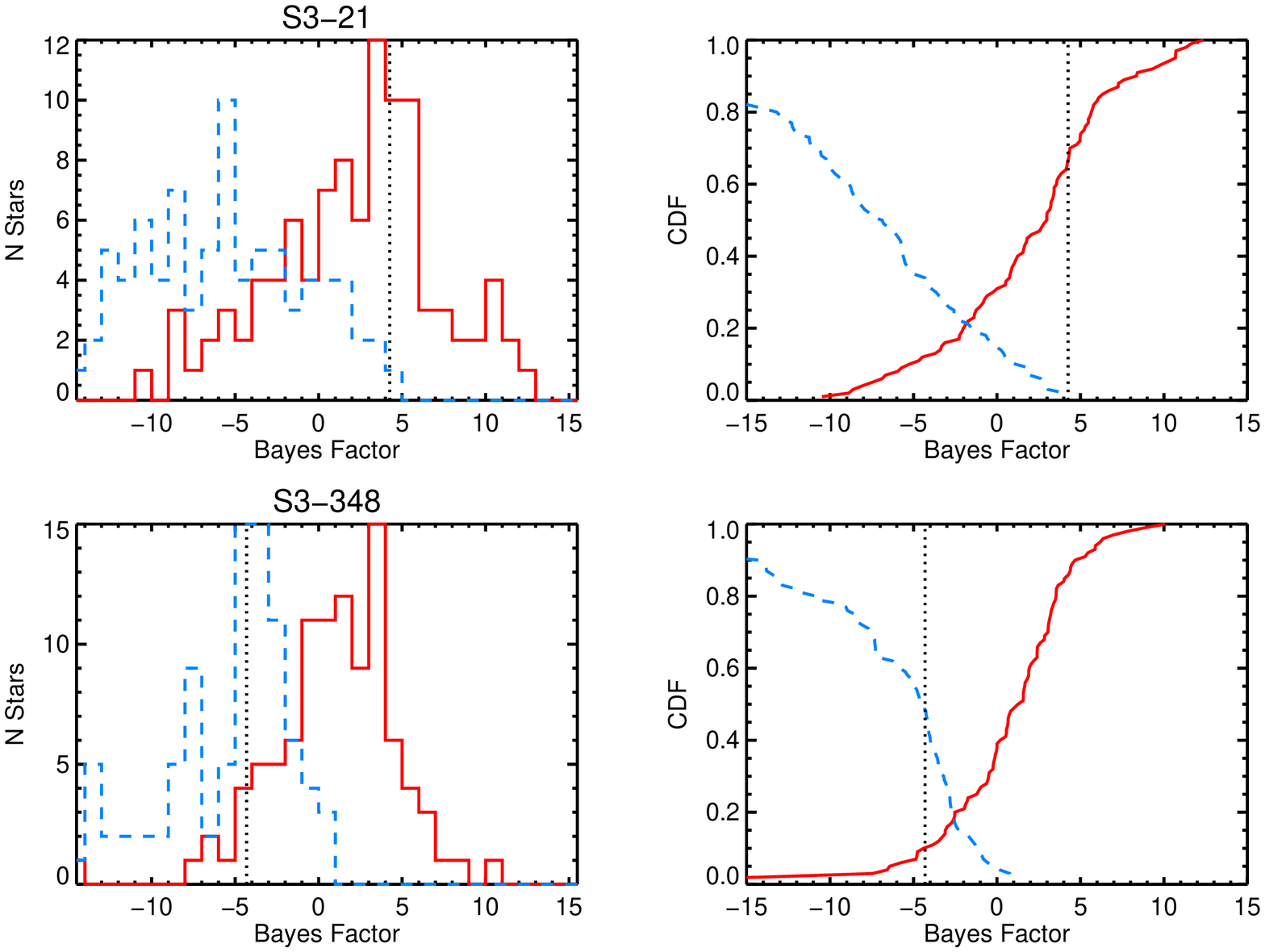}
\caption{Examples of the results of star planting simulations for stars with a high probability of being late-type (Top, S3-21) or early-type (Bottom, S3-348) are shown with their distribution of Bayes factors (we use the $\ln$ BF) measured from simulated sources planted near these stars. The plots show how the distribution of measured Bayes factor for the planted early-type stars (blue, dashed line) are separated from those of planted late-type stars (red, solid line). These distributions allow us to calibrate the measured Bayes factor (black, dotted line) for each source by sampling how likely it would be to observe a given Bayes factor if the star is early-type compared to being late-type. }
\label{fig:sim_bf_dist}
\end{figure*}

For each untyped star in our sample, the probability that it is early-type, $P_{E}(BF)$, or late-type, $P_{L}(BF)$, given the measured Bayes Factor, is: 
\begin{eqnarray}
P_E(BF) = \frac{f_{E}\Pi_{E}}{f_{E}\Pi_{E} + f_{L}\Pi_{L}}  \label{eqn:prob_early} \\
P_L(BF) = \frac{f_{L}\Pi_{L}}{f_{E}\Pi_{E} + f_{L}\Pi_{L}}. \label{eqn:prob_late}
\end{eqnarray}
where $f_E$ is the fraction of simulated young stars with Bayes Factors \textit{greater} than the measured Bayes Factor of the untyped source, and $f_L$ is the fraction of simulated old stars with Bayes Factors \textit{less} than the measured Bayes Factor of the untyped source. Figure \ref{fig:sim_bf_dist} shows an example of how $f_{E}$ and $f_{L}$ are derived. $\Pi_{L}$ and $\Pi_{E}$ are the prior probabilities of observing a late-type or early-type star, respectively (see Section \ref{sec:priors} for derivation of these priors). Statistical uncertainties on the probabilities are typically $<5$\%, assuming Poisson errors in $f_E$ and $f_L$ and errors in the priors. 

\begin{figure}[ht]
\center
\includegraphics[angle=90,width=3.5in]{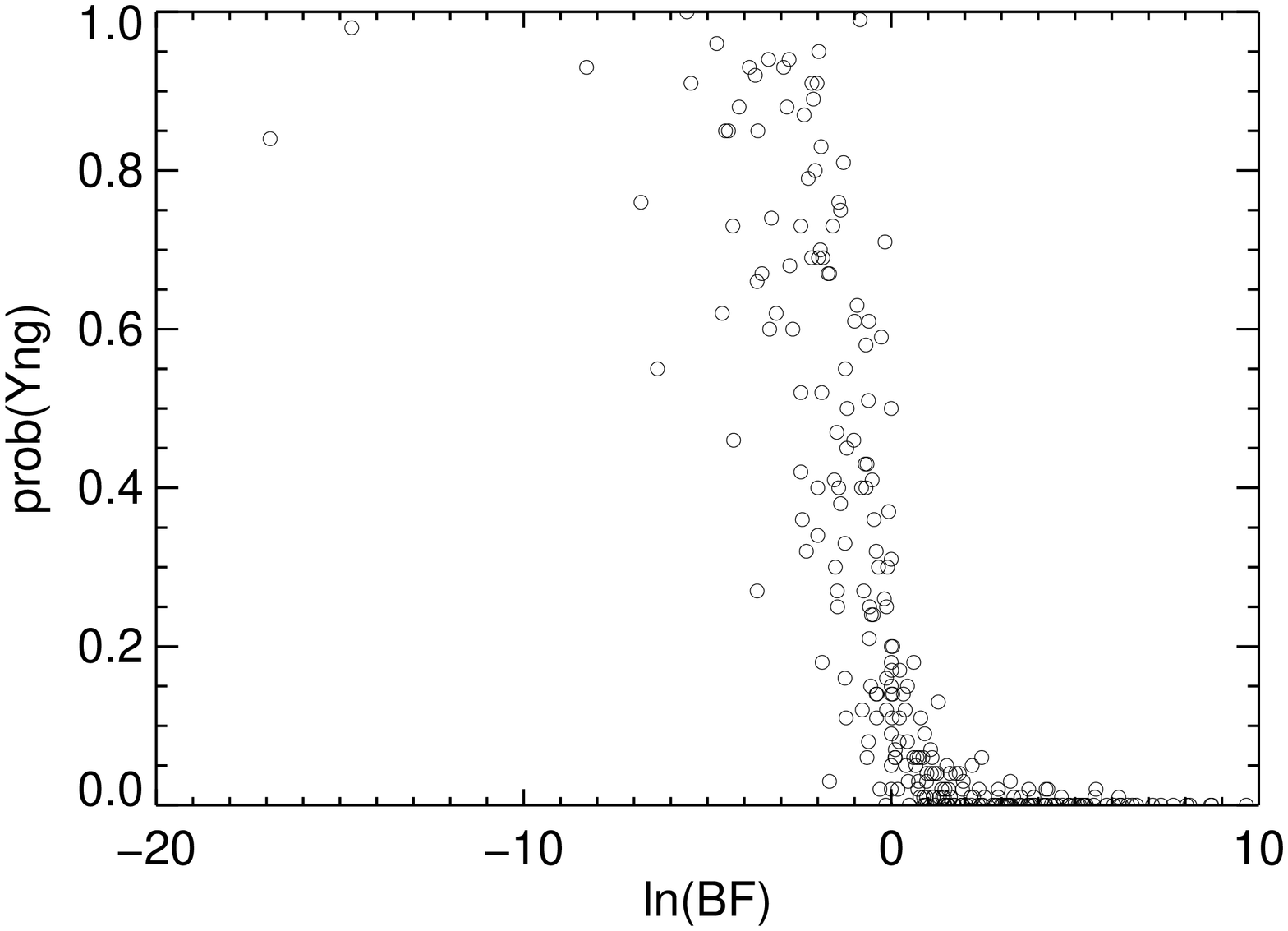}
\caption{The relationship between measured Bayes factor ($\ln BF$), and the probability that the star is young, for the set of stars without manual spectral-types. This relationship is not perfectly correlated because there is environmental variability between the location of different stars as well as differences in priors on the spectral-types.}
\label{fig:bf_v_prob}
\end{figure}

In Table \ref{tab:sim_results}, we present the location and properties of the untyped sources, as well as their associated Bayes Factor and $P_{E}$ and $P_{L}$ values from equations \ref{eqn:prob_early} and \ref{eqn:prob_late}. Figure \ref{fig:bf_v_prob} shows the probability of being an early-type star as a function of $\ln BF$ for the sources with manual spectral types as well as the untyped sources.

\subsubsection{The choice of priors}
\label{sec:priors}

A key component to assigning probabilities are the prior assumptions on the intrinsic distribution of early and late-type sources within the untyped sample. In this case, the prior is the relative probability that a star is early or late-type. 
One choice for the priors would be to assign equal probability to a star being early-type ($\Pi_E$) or late-type ($\Pi_L$) to each untyped source. Another choice would be to assume that the untyped stars have the same relative fraction of early and late-type stars as those already spectral-typed in a given magnitude bin.  Neither of these choices is entirely satisfactory; the first choice assumes that we have no information about the sources that have not been spectral-typed.
The second assumes that our sensitivity to the two types of stars is the same and that there are no location or magnitude dependent systematics. For example, in general it is more difficult to detect Br $\gamma$ than Na I at faint magnitudes given the smaller Br $\gamma$ equivalent widths and complications of background subtraction.  Additionally, the early and late-type stars have radial density profiles with very different slopes \citep{2009AA...499..483B,2009ApJ...691.1021D,2010ApJ...708..834B}. To account for this we factor our priors into two terms,
\begin{eqnarray}
\Pi_E = \Pi_{E,sens} \; \Pi_{E,R} \\
\Pi_L = \Pi_{L,sens} \; \Pi_{L,R}.
\end{eqnarray}
The first contains the differences in line sensitivity to Na I and Br $\gamma$, and the second incorporates a radial dependence in the relative number of early-type to late-type stars.

\subsubsection{The line sensitivity prior}
\begin{figure}[t]
\center
\includegraphics[width=3.5in]{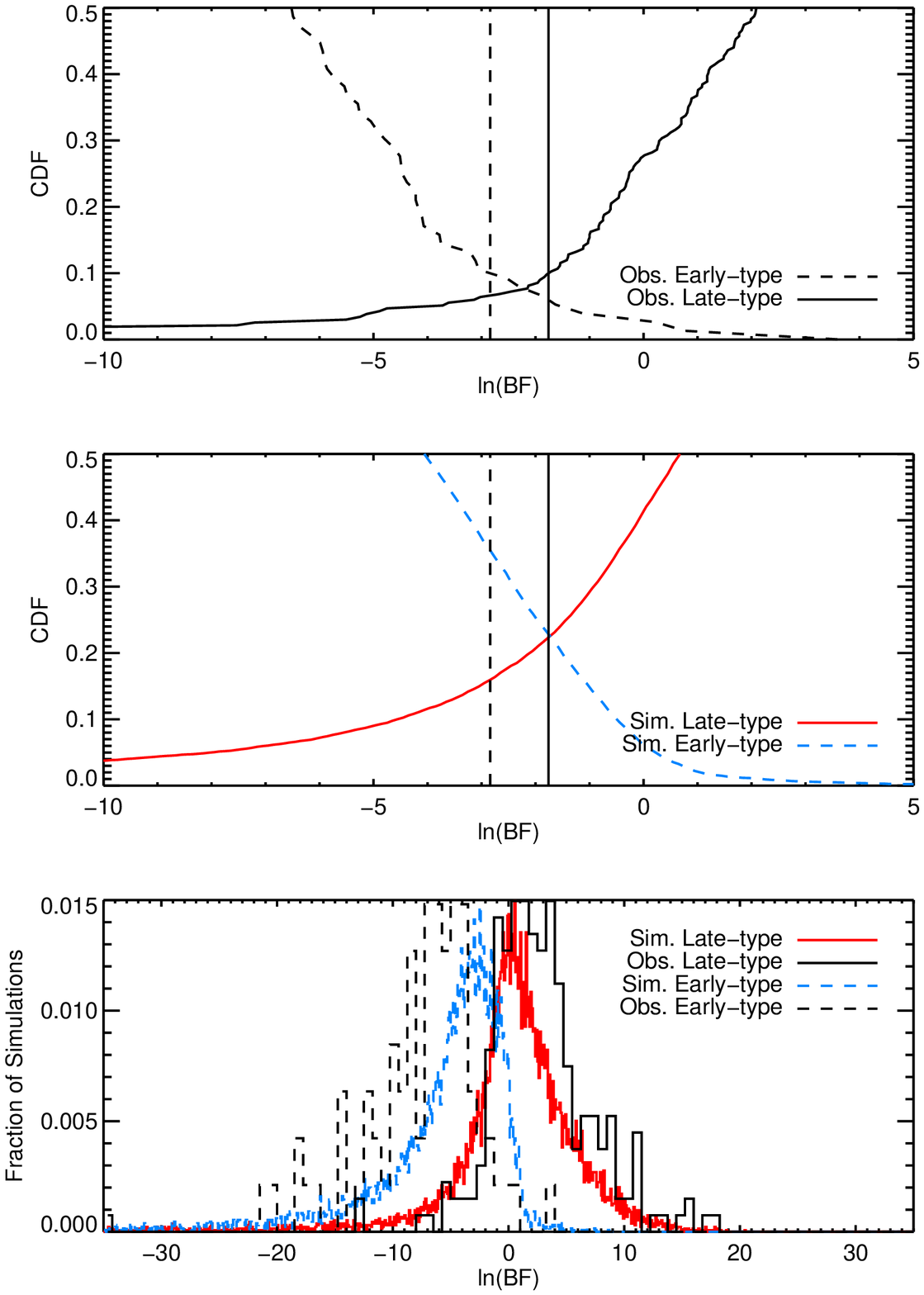}
\caption{\textbf{Top:} The choice of Bayes Factor threshold for defining early-type and late-type stars is based on the cumulative distribution of Bayes Factor for the manually classified late-type ({\em solid}) and early-type ({\em dotted}) sources. We chose the Bayes Factor threshold to be $\ln (BF) > -1.76$ ({\em solid vertical}) in order to be declared a late-type source and $\ln (BF) < -2.84$ ({\em dotted vertical}) to be declared an early-type source. These thresholds are chosen such that 90\% of the late-type stars have $\ln(BF)$ above the late-type threshold and 90\% of the early-type stars are below the early-type threshold in the manually spectral-typed sample. Note that 1 - CDF of the early-type stars (blue) is plotted to more easily illustrate how the two populations are separated. \textbf{Middle}: The simulated cumulative distributions are also shown for early-type ({\em blue}) and late-type ({\em red}) stars. The relative fractions of simulated sources that fall outside the thresholds are 60\% for early-type stars and 40\% for late-type stars. \textbf{Bottom:} The distribution of manually-typed early and late-type Bayes factors along with the corresponding simulated sources.}

\label{fig:prior_dist}
\end{figure}

To determine the relative sensitivity to the two types of stars, we turn to the results of the star planting simulations. We must set a threshold in Bayes Factor for calling a star early-type versus late-type. In this way, we can then examine the simulations to determine what fraction should have been detected as either type of star. Based on the distribution of Bayes Factor for the manually spectral-typed sources, we chose the Bayes Factor threshold to be $\ln (BF) > -1.76$ in order to be declared a late-type source and $\ln (BF) < -2.84$ to be declared an early-type source. These thresholds are chosen such that 90\% of the late-type stars have $\ln(BF)$ above the late-type threshold and 90\% of the early-type stars are below the early-type threshold in the manually spectral-typed sample (Figure \ref{fig:prior_dist}). Using these thresholds, we examine the entire set of simulated stars and compare the BF distributions with the thresholds. If we were equally sensitive to early-type and late-type stars ($\Pi_{E,sens} = \Pi_{L,sens} = 0.5$), the number of planted late-type and early-type stars with BF outside the threshold would be equal. While the distribution of Bayes Factor for the simulated late-type stars are very similar to those that have been manually spectral-typed, the distribution for simulated early-type stars are skewed closer to zero than the observed distribution  (Figure \ref{fig:prior_dist}). The relative fractions of undetected sources are 60\% early-type stars and 40\% late-type stars; this means the late-type stars are about 1.5 times more likely to be detected than early-type stars in the magnitude range of the untyped sources. We therefore use as our priors $\Pi_{E,sens} = 0.60$ and $\Pi_{L,sens} = 0.40$. We note that this prior is used as the starting point for the Bayes Factor analysis - the actual evidence for each untyped source also plays a role in calculating the posterior probability that a source is early or late-type through equations \ref{eqn:prob_early} and \ref{eqn:prob_late}.

\subsubsection{The radial distance prior}

In addition to the overall sensitivity of our survey to the two types of stars, we also consider the different radial profiles of the two populations. For example, a randomly selected star at a projected distance of $10\arcsec$ from Sgr A* has a much higher probability of being a late-type star compared to one that is at 1$\arcsec$. The distance prior is the fraction of early to late-type stars as a function of radius:
\begin{eqnarray}
\Pi_{R,E}(R) = \xi(R)/(1 + \xi(R)) \\
\Pi_{R,L}(R) = 1/(1+\xi(R)),
\end{eqnarray}
where, $\xi(R)$ is the ratio of the radial surface density profiles:
\begin{equation}
\xi(R) \equiv \frac{\Sigma(R)_{E}}{\Sigma(R)_{L}} = \frac{A_E R^{-\Gamma_E}}{A_L R^{-\Gamma_L}}.
\end{equation}

$\Pi_{R,E}(R)$ and $\Pi_{R,L}(R)$ are determined iteratively using the observations. Initially, we use as the priors, the surface density profiles of the manually typed sample of stars. This initial prior is not dependent on the probabilities from the simulations. We derive probabilities for the untyped stars using this prior along with the line sensitivity prior ($\Pi_{E,sens} = 0.6, \Pi_{L,sens} = 0.4$). Radial surface density profiles are then recalculated using the complete sample of stars, which we will use as our final radial distance prior. See Appendix \ref{append:bayesian} for details of the surface density profile measurements. The resulting power-law slope parameters: $A_E = 2.6$ stars arcsec$^{-2}$, $\Gamma_E = 0.86\pm0.13$, $A_L = 2.5$ stars arcsec$^{-2}$, $\Gamma_L = 0.01\pm0.14$, are used as the radial distance prior. We find that the resulting surface density profiles are relatively insensitive to whether they are measured with the iterative approach or stopping after the initial step; the difference in the prior probabilities between the two steps are less than 3\%.

\section{Results}
\label{sec:results}

\subsection{K$^\prime$ luminosity function}
\label{sec:kp_luminosity_function}

\begin{figure*}[bth]
\center
\includegraphics[angle=90,width=5.5in]{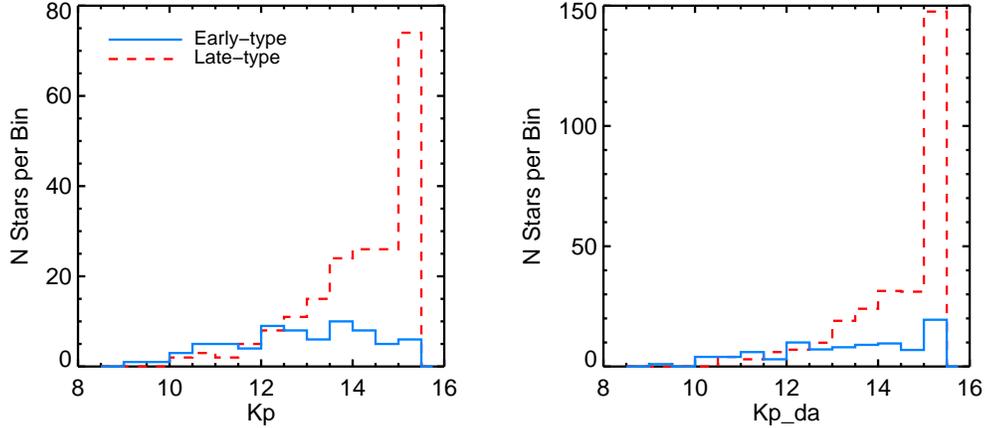}
\caption{The luminosity function of the early (blue, solid) and late-type (red, dashed) stars binned in 0.5 mag bins. \textbf{Left:} the observed luminosity function using the sample of manually typed stars and the observed $K^\prime$ magnitudes. \textbf{Right:} the differential-extinction and completeness-corrected K$^\prime_{\Delta A}$ luminosity function.}
\label{fig:kluminosity_corr}
\end{figure*}

K$^\prime$ luminosity functions are constructed for both the early and late-type stars. First, the manually-typed sample is used alone to construct these distributions by summing the number of early-type or late-type stars in each magnitude bin. We will refer to these as {\em observed} distributions, since they are equivalent to the observed KLFs reported in earlier works before correcting for incompleteness. Second, {\em completeness-corrected} KLFs are constructed by combining the manually-typed and statistically typed samples (e.g. all stars detected in our deep images) and applying a small correction for imaging incompleteness. In each magnitude bin, the number of early-type or late-type stars is given by the sum of the probabilities, divided by the imaging completeness:
\begin{eqnarray}
N_{E,mag} = \sum_i^{N_{obs}} P_{E,i} / C_{mag} \\
N_{L,mag} = \sum_i^{N_{obs}} P_{L,i} / C_{mag}
\end{eqnarray}
where $N_{obs}$ is the total number of stars, $C_{mag}$ is the imaging completeness within the given magnitude bin, and $P_{E,i}$ and $P_{L,i}$ are the the probability that the star is early or late-type, respectively. The manually spectral-typed early and late-type stars are respectively assigned either $P_{E,i} = 1$ or $P_{L,i} = 1$. Note that the imaging completeness correction is identical for both early-type and late-type samples. $N_{E,mag}$ and $N_{L,mag}$ are assumed to have Poisson errors\footnote{In cases where $N = 0$ (e.g. $\sigma_{N} = \sqrt N$), we conservatively adopt $\sigma_N = 1$.}. Poisson errors on the final number are a good approximation of the true error since we know the number of stars very well from deep images (i.e., imaging completeness is very high), and the statistical uncertainties in the probabilities for early and late-type classification are assumed to be negligible. Figure \ref{fig:kluminosity_corr} shows both the observed and completeness-corrected $K^\prime$ luminosity function for the early and late-type stars. In the faintest bin ($15.0-15.5$ mag), 96\% of the stars are detected in imaging, of which about 50\% of those have manual spectral-types. The early-type KLF increases smoothly with fainter magnitudes, while the late-type stars exhibit a large jump in the faintest bin due to the presence of red clump stars.

\subsubsection{Luminosity function of the S-stars}

\begin{figure}[bth]
\center
\includegraphics[angle=90,width=3.5in]{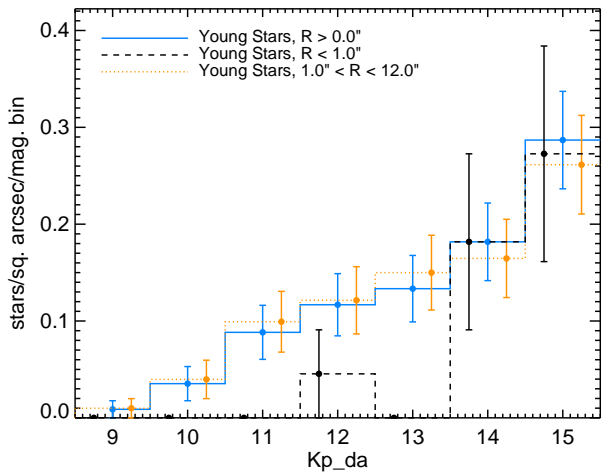}
\caption{The completeness-corrected luminosity functions of the early-type stars as a function of projected distance from Sgr A* are shown. Blue, solid - all early-type stars in the survey. Orange, dotted - early-type with projected distance of $1\arcsec < R < 12\arcsec$ from Sgr A*. Black, dashed - early-type stars within $1.0\arcsec$ of Sgr A*, scaled by a factor of 0.18 to better compare the slope of the luminosity function. The bright end ($K^\prime_{\Delta A} < 14.0$) of the luminosity function for the central $1.0\arcsec$ is inconsistent with that found further out, lacking bright stars. However, the faint end ($K^\prime_{\Delta A} > 14.0$, B-stars) of the S-stars luminosity function is consistent with the luminosity function for stars at $ > 1\arcsec$.}
\label{fig:kluminosity_compare_central}
\end{figure}

We also compare the luminosity function of the S-stars, defined here as early-type stars with $R < 1.0\arcsec$ \citep[similar to the definition in][]{2010RvMP...82.3121G}, to that of early-type stars further out. We split the sample of early-type into different regions: (1) $R < 1\arcsec$, (2) $1\arcsec < R < 12\arcsec$, and (3) all $R$. Figure \ref{fig:kluminosity_compare_central} shows the completeness-corrected luminosity functions of these three samples of stars. The stars with $R < 1\arcsec$ are 100\% complete to an extinction corrected $K^\prime_{\Delta A} < 15.5$ (11 stars: S0-2, S0-1, S0-3, S0-5,  S0-11, S0-4, S0-9, S0-31, S0-14, S1-3, S0-15). This high completeness is due to the lower than average extinction in this region as well as deeper spectroscopic observations. As observed by other studies \citep[e.g.][]{2006ApJ...643.1011P}, the central $1\arcsec$ has a lower density of stars with $K^\prime_{\Delta A} < 14.0$, compared to the outer region. However, for the stars with $K^\prime_{\Delta A} > 14.0$ in the central arcsecond, the luminosity function is statistically consistent with the faint end of the luminosity function for stars at $R > 1\arcsec$, and for the total population.

\subsection{Radial surface-density profiles}
\label{sec:radial_profile}
\begin{figure*}[thb]
\center
\includegraphics[angle=90,width=6.5in]{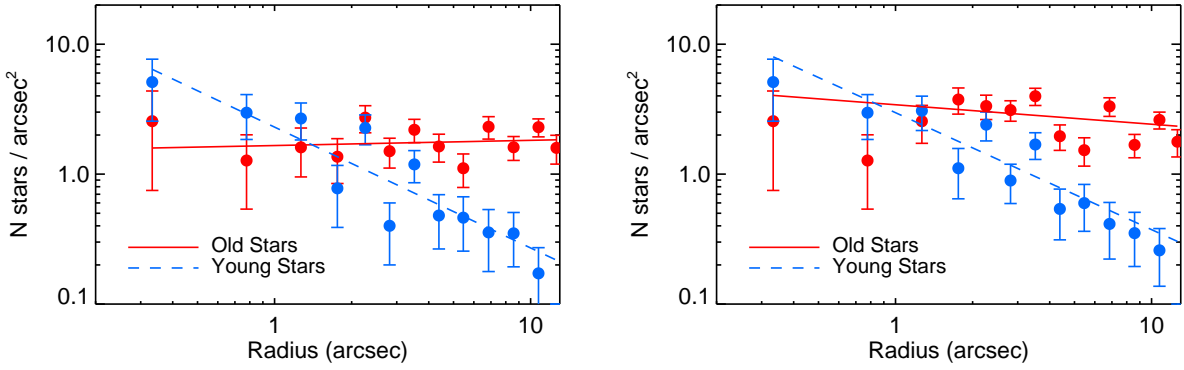}
\caption{The azimuthally averaged surface-density profile of early-type (blue) and late-type stars (red). The most probable surface-density power-law slopes are also plotted (early: dashed, late: solid). \textbf{Left:} The number counts have \textit{not} been corrected for completeness. These consist of Sample 1 stars. The most probable fit for the two populations are: $\Gamma_L = -0.04\pm0.10$, and $\Gamma_{E} = 0.93\pm0.11$. \textbf{Right}: completeness-corrected surface-density profile (Sample 4). The most probable fit for the two populations are: $\Gamma_L = 0.16\pm0.07$, and $\Gamma_{E} = 0.93\pm0.09$. For details of the completeness correction, see 
Section \ref{sec:bayesian}. }
\label{fig:radialdist}
\end{figure*}

We use a Bayesian parameter estimation method to determine the surface-density profiles of the early and late-type stars. Previous observations of the surface-density profiles of stars in this region were estimated by fitting a power-law to number counts binned by radius. The process of binning can lead to large variances in the inferred power-law fits, especially if the population is to be  separated by luminosity, for example. In order to overcome some of these limitations, we use an unbinned fit to a power-law model, $\Sigma \propto R^{-\Gamma}$. 
We compute the Bayesian posterior probability distribution for the power-law slope $\Gamma$, using the individual star's positions as the likelihood, and assuming a flat prior for $\Gamma$. See Appendix \ref{append:bayesian} for more details. The power-law slopes of the observed (as defined in Section \ref{sec:kp_luminosity_function}) surface-density profiles for the early and late-type stars is  $\Gamma_{E} = 0.93\pm0.11$ and $\Gamma_{L} = -0.04\pm0.10$, respectively. We plot these profiles in Figure \ref{fig:radialdist}, along with radially binned points for illustration and comparison. The completeness-corrected late-type surface-density profile has a best fit power-law of $\Gamma_L = 0.16\pm0.07$, while the early-type surface-density profile has a best fit slope of $\Gamma_{E} = 0.93\pm0.09$. For comparison, a traditional least-squares fit to a binned radial profile, with each radial bin containing roughly equal number of stars, has $\Gamma_{L} = 0.12\pm0.16$, and $\Gamma_{E} = 0.83\pm0.14$. While the two methods are consistent, the Bayesian method has the advantage of utilizing the precise positions of stars rather than binned positions, resulting in smaller uncertainties. Table \ref{tab:gamma} summarizes the fits to the different populations of stars.

\begin{deluxetable*}{lccccc}
\tablecolumns{6}
\tablecaption{Most probable surface-density power-law}
\tablewidth{0pc}
\tabletypesize{\footnotesize}  
\tablehead{\colhead{Population} & \colhead{completeness-corrected?} & \colhead{Radial Range} & \colhead{Magnitude Range} & \colhead{N} & \colhead{$\Gamma$}}
\startdata
Late-type & no & $> 0\arcsec$  &	all & 200 &	$-0.04\pm0.10$ \\
Late-type & yes & $> 0\arcsec$ &   all & 	305.12 & $0.16\pm0.07$ \\
Late-type & yes & $> 0\arcsec$ & $K^\prime < 14.3$ & 93.59 &	$0.27\pm0.13$ \\
Late-type & yes & $>0 \arcsec$ & $K^\prime > 14.3$ &	211.53 & $0.11\pm0.09$ \\
Early-type & no & $> 0 \arcsec$ & all &	78 &	$0.93\pm0.11$ \\
Early-type  & yes & $> 0\arcsec$ & all &	102.88 & $0.93\pm0.09$ \\
Early-type  & yes & $ > 1.0\arcsec$ & all &	89.19 &	$1.17\pm0.18$ \\
Early-type & yes & $ > 0\arcsec$ & $K^\prime < 14.3$ & 56.41	& $0.77	\pm0.13$	 \\
Early-type & yes & $ > 1.0\arcsec$ & $K^\prime < 14.3$ &	53.41 & $1.26\pm0.22$ \\
Early-type & yes & $ > 0.0\arcsec$ & $K^\prime > 14.3$ & 46.47	& $1.07\pm0.12$ \\
Early-type & yes & $ > 1.0\arcsec$ & $K^\prime > 14.3$ & 35.78 & 	$1.06\pm	0.25$ \\
Early-type & yes & $0.0\arcsec < R < 1.0\arcsec$ & $K^\prime > 14.3$ & 12 & 	$0.89\pm	0.39$ \\
Early-type & yes & $> 1.0\arcsec$ & $K^\prime < 12.25 $  & 23 & $1.51\pm 0.35$ 
\enddata
\label{tab:gamma}
\end{deluxetable*}

\begin{figure}[!th]
\center
\includegraphics[width=3.5in]{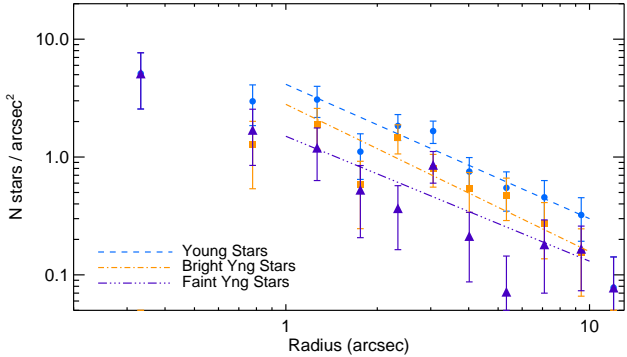}
\caption{The azimuthally averaged, completeness-corrected surface-density profile of young stars (blue). Also, the surface-density profile of those that are brighter (orange) and fainter (purple) than K$^\prime$ = 14.3, approximately the magnitude that separates B MS stars from the brighter OB supergiants and WR stars. There is a large drop in the density of bright stars in the center, but the outer radial profiles are consistent between the two populations. Power law slopes and errors for the various populations are given in Table \ref{tab:gamma}.}
\label{fig:radial_corr_yng}
\end{figure}

We also examine the early-type population for evidence of mass segregation by examining the surface-density profiles for stars brighter than K$^\prime_{\Delta A} = 14.3$, compared to the fainter population. This cut is chosen because it is approximately at the division between B-type main sequence (MS) stars and the more massive O stars and OB supergiants. This split is also motivated by the fact that there appears to be a significant lack of young stars brighter than this threshold in the central 0.\arcsec8, which has been noted by many previous observers \citep[e.g.][]{2006ApJ...643.1011P,2010ApJ...708..834B}. We wish to investigate whether there is a difference in the density structure between the B MS populations and the more massive young stars. We find that while the fainter B stars have a marginally shallower density profile, with $\Gamma_{faint} = 1.06 \pm 0.25$, than the brighter population, with $\Gamma_{bright} = 1.25 \pm 0.22$ (for $R > 1.0\arcsec$), their profiles are statistically consistent with a single power-law for projected distance R $>$ 1$\arcsec$ from Sgr A*. When including the area inside of 1\arcsec, the power-law fit for all faint B-type stars has $\Gamma = 1.06\pm0.13$, consistent with the fit for $R > 1\arcsec$. Figure \ref{fig:radial_corr_yng} shows the profiles of the different samples of stars. It is also unclear at this point whether a single power-law is a good fit to the density profiles of both populations, as there appears to be a plateau in the surface-density profile from about 1$\arcsec$ to 4$\arcsec$, beyond which the surface-density drops. This may indicate that a broken power-law is a better fit to the surface-density profile. However, because of the small number of stars, this effect is not statistically significant at this time.

\section{Discussion}
\label{sec:discussion}

\subsection{$K^\prime$ luminosity function: early-type young stars}

\begin{figure}[!th]
\center
\includegraphics[angle=90,width=3.5in]{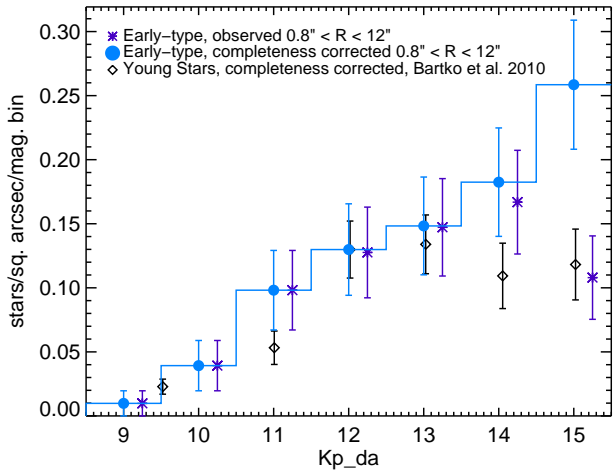}
\caption{A comparison between the $K^\prime_{\Delta A}$ luminosity function reported in this paper and that reported in the literature. The plot compares the observed (purple asterisks) and completeness-corrected $K^\prime_{\Delta A}$ luminosity function of early-type stars (blue solid) located at a projected distance $0.\arcsec 8 < R < 12\arcsec$ from Sgr A*, with the completeness-corrected K luminosity function from \citet{2010ApJ...708..834B} (black diamond) in the same radial range. The number counts are normalized between our observations and those from \citet{2010ApJ...708..834B} in the K$^\prime_{\Delta A} = 11.5-12.5$ bin in order to compare the relative difference in the slopes (the error bars are also scaled accordingly). The bin centers for the observed K$^\prime_{\Delta A}$ luminosity function are shifted slightly for clarity. We find a significantly greater fraction of faint young stars than in \citet{2010ApJ...708..834B}. The observed K$^\prime_{\Delta A}$ luminosity function is nearly identical to the \textit{completeness-corrected} luminosity function from \citet{2010ApJ...708..834B}, so any amount of incompleteness correction will lead to a greater number of faint B stars. }
\label{fig:kluminosity_compare}
\end{figure}

We find our measured luminosity function for young stars between $0.\arcsec8 < R < 12 \arcsec$ from Sgr A* to be much steeper than the one reported by \citet{2010ApJ...708..834B} in the same region. The luminosity function in \citet{2010ApJ...708..834B} is essentially flat between $K_s = 12$ and $K_s = 16$. In comparison, the luminosity function in this study rises continuously toward fainter magnitudes. Figure \ref{fig:kluminosity_compare} compares the extinction and completeness-corrected luminosity of \citet{2010ApJ...708..834B} to the present study for early-type stars at a projected distance of $0.\arcsec8 < R < 12 \arcsec$ binned in 1 magnitude bins. The luminosity functions are scaled to have the same value at $K^\prime = 12.0$ in order to compare the differences in the slopes. Even before completeness correction, our survey has a comparable fraction of young stars in the $14.5 < K^\prime < 15.5$ to that of the completeness-corrected luminosity function from \citet{2010ApJ...708..834B}. Any completeness corrections would then increase the steepness of the $K^\prime$ luminosity function in this study. The steeper slope of the luminosity function found in this study will result in a steeper inferred mass function compared to \citet{2010ApJ...708..834B}. It is not trivial to derive an IMF from a luminosity function as there are many variables that can affect the luminosity besides mass. For example, the age, star formation history, and metallicity will all affect the transformation from mass to luminosity. In \citet{lu2012}, a detailed analysis is performed using the data presented in this paper and employing a combination of stellar evolution and stellar atmosphere models. Here, we will focus our discussion on a direct comparison of the observed luminosity functions and our approach to completeness correction.

One possible explanation for the discrepancy between our KLF and previous measurements is our different method of correcting for spectroscopic completeness. Our method takes advantage of our nearly complete knowledge of the {\em location} and {\em brightness} of stars from deep imaging; only the {\em spectral-types} of some stars are unknown. We incorporate this knowledge, along with information in the spectra of these untyped stars, to assign a statistical probability of being early-type or late-type. In comparison, traditional completeness corrections ignore all information on un-typed stars and only plant simulated stars to estimate a completeness correction to be applied to the manually-typed population. Furthermore, in \citet{2010ApJ...708..834B}, the simulated stars appear to be typed in a manner that is different from the observed stars; the simulated stars are declared early-type based on CO alone, while the observed stars are declared early-type when they have Br-gamma or He I, and lack CO lines. This may lead to an over-estimate of the completeness to early-type stars, in which case, their reported numbers of early-type stars are underestimated in the faint bins where the completeness corrections dominate.

Another significant difference between this survey and that of \citet{2010ApJ...708..834B} is the region covered by the two surveys; this survey is done predominantly in the direction of the projected clockwise young stellar disk, while the survey from \citet{2010ApJ...708..834B} covers a region largely perpendicular to the disk plane (Figure \ref{fig:fields_locations}). The present survey is likely to contain more stars that belong to the clockwise disk of young stars, while \citet{2010ApJ...708..834B} showed that only one star fainter than K = 15.0 is consistent with being on this disk. A detailed kinematic analysis of the current survey is necessary to place similar constraints on disk membership (Yelda et al. in prep.), but the differences in the luminosity functions, if it is not due to differences in completeness corrections, may indicate a difference in IMF between those stars on the disk and the field population represented by the observed B stars. This would be the first indication that the stellar population is different between these two populations; previous observations of the OB supergiants and O main sequence stars have shown indistinguishable differences in number or age of those stars. A difference in the numbers of lower-mass main sequence stars could indicate differences in their formation or in their dynamical evolution. With the currently published data sets, it is not yet possible to quantitatively assess either of these scenarios. A larger systematic survey of the B stars, along with their kinematics to establish disk memberships, will be necessary to address these questions. 

\subsection{$K^\prime$ luminosity function: late-type giants}

\begin{figure}[t]
\center
\includegraphics[angle=90,width=3.5in]{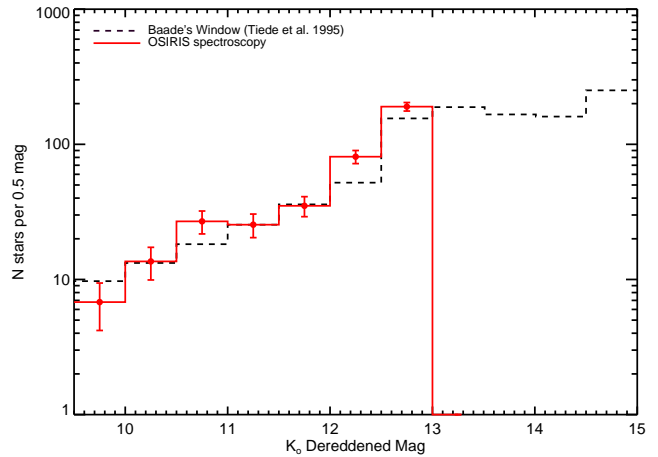}
\caption{Comparison between the completeness-corrected and dereddened luminosity function of the late-type giants observed with OSIRIS (solid, black) and the $K_o$ luminosity function observed in Baade's Window from \citet{1995AJ....110.2788T}, which gives a deep sampling of a bulge field near the Galactic center. The OSIRIS luminosity function is cut off beyond $K_o = 13.0$ (observed $K^\prime \sim 15.5$), where our completeness correction is less reliable. The amplitude of the \citet{1995AJ....110.2788T} luminosity functions are scaled to the OSIRIS luminosity function to match at the $K_o = 11.0-11.5$ magnitude bin. }
\label{fig:bw_compare}
\end{figure}

We find that the late-type luminosity function in this study is very similar to that of the inner bulge population. The late-type luminosity function is comparable to that of Baade's Window, a field located about 4$^\circ$ from the Galactic center with very low extinction ($A_K \sim 0.14$) and so has been studied extensively in the past. Figure \ref{fig:bw_compare} shows the K luminosity function from \citet{1995AJ....110.2788T}, who combined deep observations of Baade's Window with earlier work by \citet{1987ApJ...320..199F} and \citet{1993AJ....105.2121D}. These observations are corrected for reddening by dust, but not by the distance modulus ($K_o$), which reaches a depth of $K_o = 16.5$, corresponding to about $K = 19.2$ at the Galactic center when the additional extinction is included. We deredden the $K^\prime$ luminosity function from the OSIRIS spectroscopic sample in order to compare with the one from Baade's Window. We scale the amplitude of the luminosity function of Baade's Window to the $K_o = 11.0-11.5$ bin of the completeness and dereddened luminosity function. We find that the two generally agree well, down to the spectroscopic limit at $K^\prime = 15.5$, or $K_o = 12.5-13.15$, where the red clump stars are concentrated in the luminosity function. Below $K^\prime = 15.5$, we no longer have spectroscopic differentiation between the young and old population, but it is clear from Figure \ref{fig:bw_compare} that the GC luminosity function matches the red clump features from Baade's Window. The match in the slope and location of the red clump to that of the bulge indicates that the star formation history at the Galactic center may be similar \citep[see also][]{2007ApJ...669.1024M,2011ApJ...741..108P}. 

\subsection{Surface-density profile of young stars}

The power-law fit to the projected surface-density profile of all the young stars is consistent with that of previous observations ($\Gamma = 0.90\pm 0.09$). \citet{2006ApJ...643.1011P} reported a surface-density profile power-law slope of $\Gamma = 2.1 \pm 0.17$ in the plane of the young stellar disk, including only stars in the clockwise disk. As the current study does not separate the young stars into disk members and non members, we must compare the surface-density profiles for all young stars. Using the \textit{entire} sample of young stars in \citet{2006ApJ...643.1011P} within the field of view of this study, regardless of disk membership, the best fit power-law to the projected surface-density profile is $\Gamma = 1.12\pm0.13$, consistent with the measurement here. Our results are also consistent with those from \citet{2009AA...499..483B}, who used medium-band imaging to separate early-type stars with low CO equivalent widths from late-type stars with strong CO absorption at $\sim 2.3$ $\micron$; \citet{2009AA...499..483B} found that at a projected distance of 1$\arcsec$ to 10$\arcsec$ from Sgr A*, the early-type stars have $\Gamma = 1.08\pm 0.12$.  With the identification of these young stars, further insights into their origins can be obtained with the addition of kinematic data to reconstruct their orbital properties (Yelda et al. in prep.).

\subsection{Properties of the S-stars}

Our observations show that other than a deficit of stars brighter than $K^\prime_{\Delta A} < 14.0$, the early-type S-stars within the central $\sim1\arcsec$ have very similar properties to those found at greater distances. There are no significant differences in the surface-density profile of the B-type stars (stars with K$^\prime_{\Delta A} < 14.3$) within and outside the central arcsecond (Table \ref{tab:gamma}). The B-stars show a continuous surface density profile throughout the survey region. The luminosity function within the central arcsecond is also consistent with the early-type stars further out for $K^\prime_{\Delta A} > 14.0$ (Figure \ref{fig:kluminosity_compare_central}). In contrast, \citet{2010ApJ...708..834B} found that the luminosity function of stars with $R < 0.\arcsec8$ is significantly steeper than that of stars with $0.\arcsec8 < R < 12\arcsec$. For comparison, we also constructed a luminosity function for early-type stars with $R < 0.\arcsec8$. The luminosity function in this region is the same as \citet{2010ApJ...708..834B}, because of the high completeness in this region for both surveys (the spectral identifications are the same in this region). We find, as in \citep{2010ApJ...708..834B}, that this region is missing bright stars with $K^\prime_{\Delta A} < 14.0$ compared to the luminosity function at $R > 0.8\arcsec$. However, we find that the faint end of luminosity function ($K^\prime_{\Delta A} > 14.0$) inside $R < 0.\arcsec8$ is consistent with the faint end of the luminosity function  for stars with $0.\arcsec8 < R < 12\arcsec$, due to the steeper measured luminosity function in this survey. Our results suggest that the B-type S-stars may represent a continuous population of B-stars throughout the central 0.5 pc. It is unclear at this time however, whether the S-stars can have originated from the most recent star formation event that formed the young disk of stars further out; for example, an explanation for the curious deficit of bright stars in the inner $1\arcsec$ will be necessary for this hypothesis. Resolving this issue will have a strong impact on our understanding of the star formation in the region as well as the timescales for the dynamical mechanisms that are necessary to bring the S-stars so close to the supermassive black hole. 

\subsection{Cusp clearing out to 0.5 pc}

The observed flat surface-density profile of the old red giants extends out to the edge of our survey at about 0.5 pc, about a factor of 3 further than our initial spectroscopic survey in \citet{2009ApJ...703.1323D}. While this survey predominantly samples the region east of Sgr A*, it should be representative of the distribution of old stars in this region as there do not appear to be any detectable deviations from spherical symmetry. This large core profile is also consistent with the narrowband imaging results from \citet{2009AA...499..483B} and the spectroscopic results from \citet{2010ApJ...708..834B} based on samples more to the north of Sgr A*. It is unclear from the present data whether there is a break  in the surface-density profile at larger radii, as the survey truncates at about 12$\arcsec$. Using a broken power-law model, we have attempted to constrain the break radius and the outer power-law slope, but the current data have insufficient radial coverage to strongly constrain the location of the break. Because the projected surface density profile is so flat, it is also difficult to determine the true \textit{spatial} density profile of the late-type stars ($\rho(r) \propto r^{-\gamma}$, where $r$ is a 3D distance). This limitation can be removed with the inclusion of kinematic information. For example, Jeans modeling of proper motion and radial velocity measurements has successfully been used by \citet{2012JPhCS.372a2016D} to constrain the power-law exponent, $\gamma$, within the survey region presented here.

\section{Conclusions}
\label{sec:summary}

In this paper, we report the results of our new spectroscopic survey of the central 0.5 pc of the Milky Way. This study presents both new observations as well as new methodologies. Our new data extend previous spectroscopic observations along the disk plane of the young stars by a factor of $\sim2-3$. We develop a new method for statistical spectral-typing that we use for completeness correction. This method allows us to takes advantage of prior information about the stars. Most importantly, this includes information about the locations, brightnesses, and spectra. Because over 95\% of stars within the magnitude range of interest are detected in imaging, this provides a very robust method for constructing completeness-corrected luminosity functions and surface-density profiles for young, early-type stars and old, late-type giants.

We find that the measured radial surface-density profiles are consistent with previous studies \citep{2009AA...499..483B,2009ApJ...703.1323D,2010ApJ...708..834B}. The surface density profile of the late-type stars appears flat within our survey region, suggesting that the `core' in the red giants is at least $\sim 0.5$ pc in size. The early-type stars have a much steeper radial surface density profile such that they dominate the stellar density within $\lesssim 0.04$ pc from Sgr A*. 

The luminosity functions of both the late and early-type stars rise towards fainter magnitude bins. The late-type stellar luminosity function is consistent with the inner bulge of the Galaxy, indicating that the star formation history of the Galactic center may be similar. The luminosity function for the early-type stars is consistent with that of previous studies at the bright end \citep[K$^\prime < 13.0$;][]{2006ApJ...643.1011P,2009AA...499..483B}, but is steeper than reported in a recent study at the faint end \citep{2010ApJ...708..834B}. This steepening of the faint end of the luminosity function will likely result in a steeper mass function than the very top-heavy IMF reported in \citet{2010ApJ...708..834B}. The derivation of a mass function from a luminosity function is presented in Paper II \citep{lu2012}. 

We find that the S-stars at $R < 1\arcsec$ and $K^\prime_{\Delta A} > 14.0$ (B-type) have the same luminosity function and surface density profile as the B-type stars further out. This suggests that the population of all B-type stars in the central 0.5 pc may be related, though there is insufficient information at this time to determine whether they originate from the same star formation event as the young stellar disk \citep[see also, Paper II,][]{lu2012}. 

Accurate measurements of the luminosity function are important, as different luminosity functions can lead to very different conclusions about star formation in the extreme tidal environment of the Galactic center, which affects our understanding of star formation in general. It is important to note that current spectroscopic studies are limited to $K^\prime < 15.5$ mag, which corresponds to $\sim 10 M_\odot$ \citep[Paper II][]{lu2012}. To more completely compare the Galactic center mass function with local star-forming regions, observations down to about a solar mass are necessary. This mass corresponds to about $K^\prime \approx 21$ mag at the Galactic center, which cannot be reached by current IFU instruments. This regime for scientific study will only be opened with future Giant Segmented Mirror Telescopes (GSMT), such as the Thirty Meter Telescope (TMT).

The authors thank the staff of the Keck observatory, especially Randy Campbell, Al Conrad, and Jim Lyke, for all their help in obtaining the new observations, Rainer Sch{\"o}del for providing us with an extinction map of the Galactic center region, and Annika Peter for discussions about the Bayes factor. This work has been supported by  a TMT postdoctoral fellowship (T. Do), a NSF Astronomy \& Astrophysics Postdoctoral Fellowship (AST-1102791; J. R. Lu),  NSF grant AST-0909218 (PI Ghez), and the Levine-Leichtman family foundation. The infrared data presented herein were obtained at the W. M. Keck Observatory, which is operated as a scientific partnership among the California Institute of Technology, the University of California and the National Aeronautics and Space Administration. The Observatory was made possible by the generous financial support of the W. M. Keck Foundation. The authors wish to recognize and acknowledge the very significant cultural role that the summit of Mauna Kea has always had within the indigenous Hawaiian community. We are most fortunate to have the opportunity to conduct observations from this mountain.

{\it Facilities:} \facility{Keck:II (OSIRIS)}, \facility{Keck:II (NIRC2)}.


\input{ms.bbl}
\clearpage
\appendix

\section{SPECTRAL-TYPE VERIFICATION}

\label{sec:verification}
\begin{figure}[ht]
\center
\includegraphics[angle=90,width=7in]{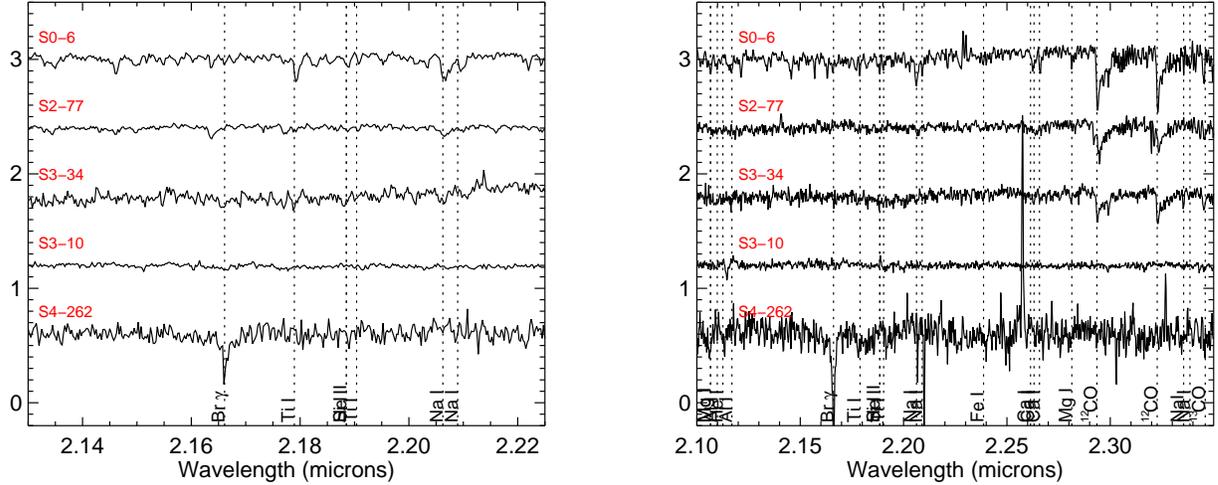}
\caption{\textbf{Left:} Spectra of sample sources in the Kn3 filter illustrating the types of stars seen in our sample. From top to bottom, the spectra are: 1) a typical late-type stellar spectrum with prominent Na I lines at 2.206 and 2.2090 \micron; 2) and 3) are late-type stars with small Na I equivalent widths; 4) an early-type star with a featureless spectrum in Kn3; and 5) a typical early-type star with Br $\gamma$ absorption at 2.1661 \micron. \textbf{Right:} The K broad-band spectra for the same sources, showing how the CO band-head can be used as discriminator of early-type versus late-type for the more ambiguous sources in Kn3. Featureless early-type sources in Kn3, such as S3-10, may also have He absorption in the broad-band filter.}
\label{fig:spec_verification}
\end{figure}

The identification of spectral types is largely based on the presence or absence of the Br $\gamma$ and Na I lines. The are, however, a few bright sources for which there are no spectral features or very weak Na I lines. In \citet{2009ApJ...703.1323D}, we classified the bright, $K^\prime < 14$ featureless stars within the Kn3 wavelength range as young, as they were presumed to be OB stars with very low or zero equivalent width in Br $\gamma$, consistent with the variance in Br $\gamma$ reported by \citet{1996ApJS..107..281H}. However, at the faint end of this range (K$^\prime$ = 13-14), we have determined that a few sources contain very weak Na I features, with equivalent widths $\lesssim 2$ \AA, compared to the average Na I equivalent width of $\sim 3.8$ \AA. In order to determine whether these sources, and other similar ambiguous sources, are young or old, we obtained K-broad band spectra with OSIRIS for a sample of stars having Na I equivalent width $< 2$ \AA (Table \ref{tab:change_spectral_type}). Figure \ref{fig:spec_verification} shows examples of the low Na equivalent width sources compared to more typical early-type sources, and the corresponding spectra in the K broad-band filters. The K broad-band spectra cover the CO band-heads at 2.3 $\micron$, which are very strong in late-type stars and are much better discriminators of the temperature of the star. We find that the sources with low, but detectable, Na I equivalent widths in the Kn3 filter have detectable CO features, indicating that they are late-type stars. However, the CO equivalent widths are smaller than the majority of the K and M giants. This is consistent with those of warmer giants of $\sim 5000$ K instead of  with temperatures in the range $3000-4000$ K \citep{2000AJ....120.2089F} for K and M stars. These stars likely represent a younger population (100-300 Myr old) compared to the $\sim 1$ Gyr old M and K giants, but we will classify them as late-type to separate them from the much younger $\sim 6$ Myr old population \citep[see also][]{2003ApJ...597..323B,2011ApJ...741..108P}. Based on these observations, we make two minor modifications for the spectral-type classification criteria: (1) the brightness cut off for featureless stars to be classified as early-type is moved from K$^\prime < 14.0$ to K$^\prime < 13.0$, (2) sources with small, detectable Na I equivalent widths are classified as late-type. 
This results in the reassignment of the spectral types of 8 stars from \citet{2009ApJ...703.1323D} from early-type to late-type, 4 of which have $K^\prime > 15.0$. We find that these modifications have resulted in a much more robust method for spectral classification, with no incorrect assignments when comparing the followup sample between Kn3 and Kbb. We include the properties of stars classified using Kbb spectra in the Bayesian inference model through the distribution of equivalent widths in Br $\gamma$ and Na I, as measured using the Kn3 spectra (Section \ref{sec:bayesian}).

\begin{deluxetable*}{lccccc}
\tablecolumns{6}
\tablecaption{Stars observed in K broad-band for spectral-type verification}
\tablewidth{0pc}
\tabletypesize{\scriptsize}  
\tablehead{\colhead{Name} & \colhead{K$^\prime$} & \colhead{ln(BF)} & \colhead{Kn3 Sp. Type\tablenotemark{a}} & \colhead{Kbb Sp. Type} & \colhead{Field}}
\startdata
\input{verification_sources}
\enddata
\tablenotetext{a}{Initial spectral-type using only Kn3 spectra with the criteria described in this paper in Section \ref{sec:manual_spec}.}
\label{tab:change_spectral_type}
\end{deluxetable*}

\section{MEASUREMENT OF SPECTRAL FEATURES}
\label{sec:auto_spec_type}
\begin{figure}[ht]
\center
\includegraphics[angle=90,width=6.0in]{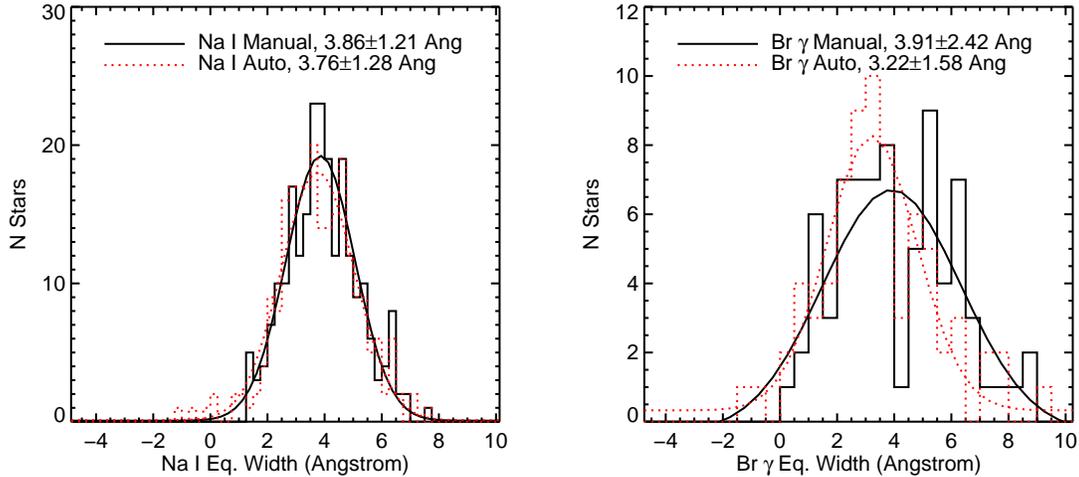}
\caption{\textbf{Left:} the observed distribution of Na I equivalent widths (solid black) as measured manually compared to that using the automated procedure described in Section \ref{sec:measurement} for sources with $K^\prime > 14.0$. \textbf{Right:} a similar plot for the distribution of equivalent widths for Br $\gamma$ at 2.1661 $\micron$. These distributions show that the automated algorithm produces similar results to measurements that require human interaction.}
\label{fig:na_br_eqwidth_dist}
\end{figure}

In order to facilitate source extraction and identification, and to run Monte Carlo simulations for completeness corrections, we also develop a method for automating the measurement of the equivalent widths of the Br $\gamma$ and Na I doublet lines as well as the radial velocity of the sources. The steps in this procedure are as follows:
\begin{enumerate}
\item We cross-correlate the spectrum with that of M3II giant HD40239 from the SPEX telescope infrared spectral templates \citep{2009ApJS..185..289R} in the range of the observed filter from 2.121 to 2.220 \micron; this template was chosen because it has a high peak correlation value when cross-correlated with most of the observed late-type sources.
\item If the correlation coefficient is greater that 0.5, then the radial velocity is measured based on the peak of the cross-correlation function. We determine the thresholds for correlation from the $K^\prime > 14.0$ subset of manually spectral-typed sources (Sample 2). The location of the cross-correlation peak is determined by fitting a parabola to the five closest points around the pixel with the maximum correlation. Once the radial velocity is determined, we then shift the spectrum to rest wavelengths and measure the equivalent width of the Na I doublet. The doublet is measured by integrating the continuum-removed spectrum between 2.2053 to 2.2101 $\micron$ as in \citet{2000AJ....120.2089F}. 
\item If the cross-correlation coefficient is below 0.5, the wavelength range over which we performed the cross correlation around the Na I doublet is reduced to between 2.20 to 2.215 $\micron$. By restricting the wavelength range, the sensitivity to the Na I feature is increased for low SNR spectra. If the peak of the cross-correlation function is now greater than 0.5, the radial velocity and the Na I equivalent width is measured in the same way as in step 2. If the correlation peak is still less than 0.5, we do not apply any velocity shifts to the spectrum, but still integrate over the region around Na I doublet to establish the equivalent width within this region. 
\item We then measure the equivalent width around Br $\gamma$ by first cross-correlating the spectrum at $\pm 4000$ km/s around the Br $\gamma$ line with a template spectrum constructed from multiple observations of S0-2 and shifted to rest wavelengths. If the cross-correlation coefficient is greater than 0.3, we fit a Gaussian to the region $\pm 0.1$ $\micron$ around the wavelength corresponding to the peak lag in the cross-correlation function. If the peak correlation is less than 0.3, we fit a Gaussian to the wavelength region $2.1661 \pm 0.1$ $\micron$. Because we are interested in measuring the faint B stars, which should have their Br $\gamma$ in absorption, we restrict the Gaussian fit to absorption features. We also require that the width of the Gaussian fit be greater than 1 spectral pixel in order to avoid fitting cosmic rays or bad detector pixels. 
\item The errors on the measured equivalent widths and radial velocities are estimated by splitting the data into three subsets. In order to obtain comparable SNR between these three spectra, we sort all of the spectra for each star by their SNR and populate each subset such that the resulting combined spectra would have similar SNR. We then apply steps 1-4 to each of the three spectra. We use the standard deviations of the radial velocity and equivalent widths as the error for the corresponding measurement. Most stars have between 6 and 9 spectra observed, though there are stars with as few as one measurement if it is at the edge of our dither pattern. For stars with fewer than three measurements, we estimate the error using a fit for the correlation between SNR and equivalent width uncertainties. Using the power-law fit to this relationship and the given SNR of the spectrum, we infer the error on the measured parameters.  
\item To filter out spurious fits, we do not consider stars with Na I or Br $\gamma$ equivalent widths $> 15$ \AA. This threshold is set above what is physically expected for any star at the Galactic center \citep{1996ApJS..107..281H,2000AJ....120.2089F}. Stars with spurious detections are flagged and those measurements are not considered in the subsequent analyses. Approximately 13\% of the stars in our sample of stars with no manual spectral types (Sample 3) have either equivalent width measurements above this threshold using the automated routine, or the routine was unable to provide an equivalent width measurement. These stars tend to have have SNR $< 5$.
\end{enumerate}
Because these measurements are fully automated, they may be susceptible to systematic errors in the spectra that would lead to poor estimates of the equivalent widths of the spectral lines. To test the accuracy of this automated method, we compare the results to those of measurements that have been individually extracted and checked by eye. For the late-type giants, we find that the measurements using the automated routine are consistent with manual measurements; the mean of the distribution of Na I equivalent width for old stars with brightness $K^\prime$ $<$ 14.0 is $3.8 \pm 1.3$ \AA~ compared to $3.9 \pm 1.2$ \AA~ measured manually. For the early-type stars, we find the mean equivalent width of Br $\gamma$ for young stars with K$^\prime$ $<$ 14.0 from the automated routines ($3.2 \pm 1.6$ \AA~) to be consistent with the equivalent widths measured manually ($3.9 \pm 2.4$ \AA). See Figure \ref{fig:na_br_eqwidth_dist} for plots of the distributions of equivalent widths of Br $\gamma$ and Na I. We conclude that the automated and manual procedures achieve comparable measurements.

\section{PHOTOMETRIC CALIBRATION}
\label{sec:photo_calib}
\begin{figure}
\begin{center}
\includegraphics[width=4in]{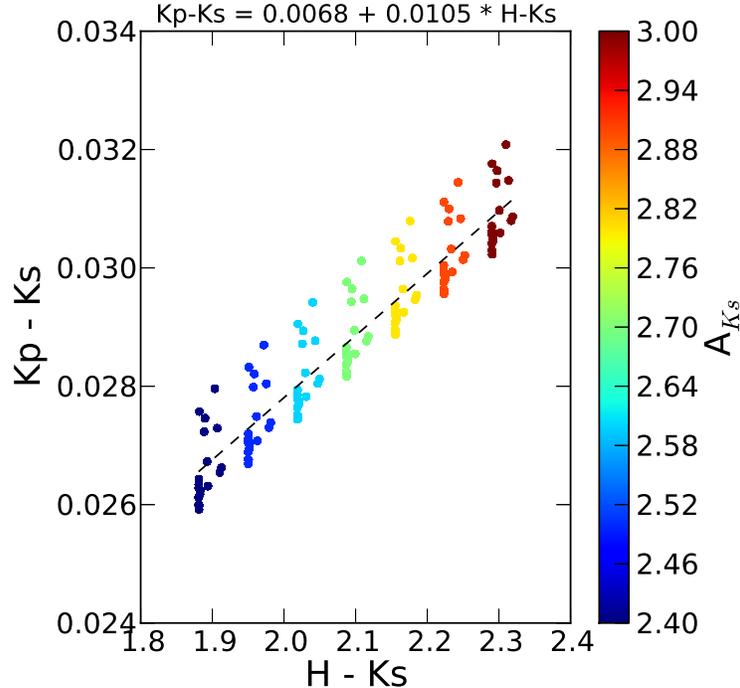}
\end{center}
\caption{The K$^\prime$ - Ks correction as a function of H - Ks color derived from synthetic photometry of red giant stars observable at the Galactic center. 
Synthetic stars are selected from a simulated 5 Gyr starburst with solar metallicity
at a distance of 8 kpc and reddened using the \citet{2009ApJ...696.1407N} Galactic center extinction law.
Stars on the red-giant branch and in the red-clump are the most numerous in Galactic center
observations, which corresponds to selecting stars from the synthetic isochrone
that have $14<$ Ks $<17$ at A$_{Ks}=2.7$.
These stars are shown for extinctions ranging from $2.4 < $ A$_{Ks} < 3.0$, the observed
range of extinctions in the central parsec of the Galaxy. A linear relation between
K$^\prime$ - Ks and H - Ks is fit and used to convert between Ks magnitudes reported in \citet{2010AA...511A..18S} and the K$^\prime$ magnitudes reported in this paper.
}
\label{fig:filter_conversion}
\end{figure}

Starlists are photometrically calibrated using published magnitudes 
reported in \citet{2010AA...511A..18S}. 
They provide an extensive star list with absolute H, Ks, and L' photometry and we select calibrator stars as those with brightnesses of H$<$18, Ks$<$16, L'$<$15,
and photometric errors of $\sigma_H<0.03$, $\sigma_{Ks}<0.02$, and $\sigma_{L'}<0.04$.
We also exclude stars that have neighbors within 0\farcs3 and $\Delta Ks<-1.5$. 
The Ks magnitudes are then converted into our K$^\prime$ filter set using the equation
\begin{equation}
K^\prime = Ks + 0.00683 + 0.01049 * (H - Ks)
\end{equation}
This K$^\prime$$-$Ks conversion equation is determined by simulating synthetic 
spectra for a stellar population with an age of 5 Gyr at a distance of 8 kpc
using models of stellar evolution and atmospheres described in \citep{lu2012}.
The synthetic spectra are reddened using the Galactic center extinction law by 
\citet{2009ApJ...696.1407N} and a range of extinction values ($A_{Ks}=2.4$-$3.0, \Delta A_{Ks}=0.1$).
The reddened synthetic spectra are convolved with atmospheric and filter transmission profiles using the package {\em pysynphot} to generate synthetic photometry for a suite of near-infrared filters, including H, K$^\prime$, and Ks. 
A linear relation is then derived for the subset of simulated stars that fall along the red giant branch from Ks=14-17, since most of the 
observed calibrators are cool, red giants (Figure \ref{fig:filter_conversion}. 
Simulations of a younger (6 Myr) population
shows that the K$^\prime$$-$Ks conversion for hot, young stars would result in K$^\prime$ photometric differences of less than 0.01 mag. 
This is far less than the 0.06 mag zeropoint error in the calibrator Ks magnitudes \citep{2010AA...511A..18S}. 
The final photometric errors for the sample of stars that are brighter K$^\prime$=15.5 has a mean of 0.08 mag.

\subsection{IMAGING COMPLETENESS}
\label{sec:image_completeness}

The NIRC2 imaging completeness as a function of position and brightness is estimated 
by planting simulated stars and determining how well they can be recovered. 
To accurately estimate the completeness, the images containing simulated stars must
be analyzed in exactly the same fashion as the real images. 
We therefore plant stars in each tile of the mosaic as well as each tile's three subset 
images at the same position and brightness. 
The simulated images for each tile are then analyzed with the same requirement that 
sources must be detected in both the combined tile image and all three of its 
subset-images to be identified as a star.
It is not necessary to plant stars at the same position in multiples tiles where 
the tiles overlap as the mosaic process did not impose additional requirements for sources 
to be detected in more than one tile. 

Artificial stars are generated in a grid in both magnitude and position. 
Simulated magnitudes range from K$^\prime$=7.7 $-$ 19.5 in steps of $\Delta$K$^\prime$ = 0.25 mag. 
Simulated positions are set in a regular grid separated by 0\farcs25. 
This grid of artificial stars cannot be planted in a single simulated image without dramatically impacting the stellar density and resulting completeness measurements. 
Therefore many simulated images are generated, each one containing artificial stars at a fixed brightness and spaced 0\farcs5 apart \citep[see also,][]{2007AA...469..125S}. 
For a given brightness, 4 simulated images are required to achieve the final 0\farcs25 
spatial sampling. 
Over 300 simulated images are necessary to cover the full brightness range just for
the combined image of each tile.
In total, 16,784 simulated images were generated for the 13 tiles in 2006, including their
total and three subset images, and were analyzed using Starfinder in an identical 
manner to the observed images. 
This required $\sim$1400 computing hours, or about 1 week when run in parallel on several powerful desktop computers. 
The simulated and recovered starlists for all the tiles were mosaicked together in the same 
manner as the observed starlists.
The final outcome of the star planting simulations is a 3D cube of the number of simulated
and recovered stars at different X, Y, and K$^\prime$ values, allowing completeness curves to be 
calculated for different areas of the image. 
Star planting simulations are extremely time consuming; therefore completeness maps are
constructed only for a single epoch, 2006, which has slightly higher Strehls than the other two epochs. 
This may tend to overestimate the completeness; however, the effect is negligible at K$^\prime$$<$16 where luminosity functions are analyzed in this paper.
In this work, the completeness is calculated for the entire OSIRIS field of view.
Figure \ref{fig:completeness} shows the resulting completeness curve and how completeness changes with radius.
Completeness decreases inside of 4'', primarily due to the limited contrast around 
the bright Wolf-Rayet stars concentrated in this central region around Sgr A*.

\section{OSIRIS STAR PLANTING SIMULATIONS}
\label{sec:sims}

Our effort to spectral-type the stars using the OSIRIS Kn3 data are subject to several limitations, which results in incompleteness in our survey for stars with K$^\prime$ $> 12.5$ mag. The following is a list of contributors to the incompleteness of the survey:
\begin{itemize}
\item Intrinsic variations in the equivalent widths of absorption lines will make some stars more difficult to detect than others at a given SNR.
\item The halo noise from nearby bright stars adds both photon noise and possibly systematic errors in some absorption features. For example, we are unable to obtain reliable spectra for bright stars  within $\sim0.\arcsec25$ of the WR stars.
\item Photon noise from the background as well as from dark current are significant sources of noise for isolated stars. 
\item Spatially varying background, especially emission from Br $\gamma$ gas at the Galactic center may lead to systematic errors in the equivalent width measurements around Br $\gamma$.
\item Read noise from the detector also contributes to the noise, but its contribution is insignificant compared to the other sources of error.
\end{itemize}
Some of theses sources of uncertainty affect the early-type and late-type stars differently; for example, the background of Br $\gamma$ emission may contribute a systematic bias to the intrinsic Br $\gamma$ absorption line for young stars, but will not affect the measurement of equivalent widths of the Na I doublet. However, the late-type stars will not entirely escape this effect because the measurement of the equivalent width of the wavelengths near Br $\gamma$ is a parameter in the Bayesian evidence. This error is very spatially dependent, as it is the result of a poor estimate of the background at the location of the star. The most problematic regions are therefore regions where there are strong spatially varying Br $\gamma$ emission, such as in the Mini-Spiral \citep[e.g.][]{2004AA...426...81P}. 

\begin{figure}[bth]
\center
\includegraphics[width=3.5in]{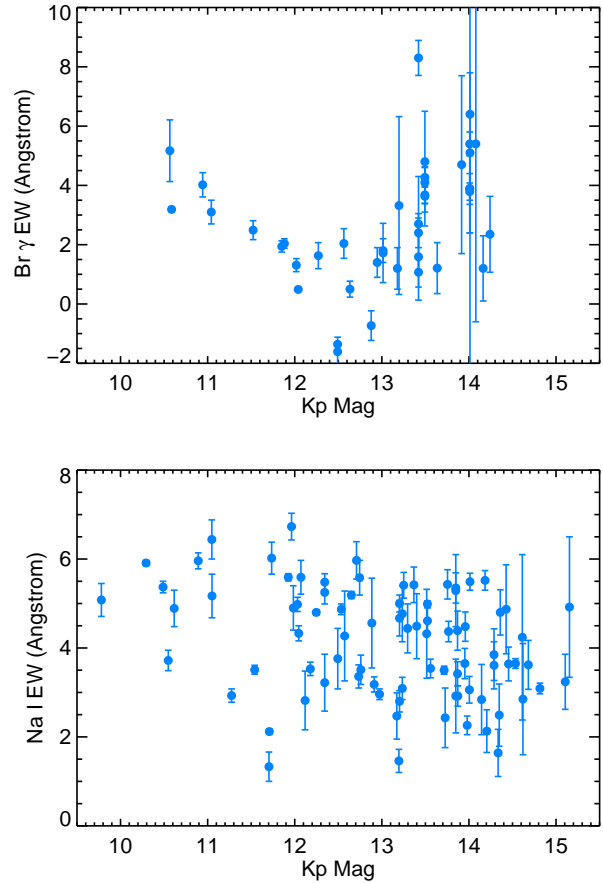}
\caption{The equivalent width of Br $\gamma$ (top) and Na I (bottom) for the sample of stars used for the star planting simulations as a function of $K^\prime$ magnitude.}
\label{fig:star_planting_ew}
\end{figure}

\begin{figure}[bth]
\center
\includegraphics[width=3.5in]{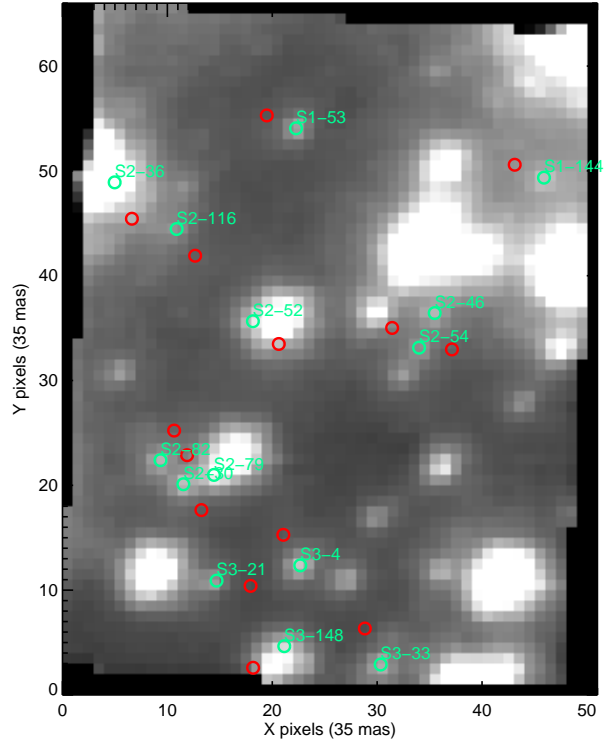}
\caption{Example of a star planting simulation. This example is taken from the field GC East (PA = 285 deg), shown before star planting simulations. Labeled with green circles are stars with $K^\prime < 16.0$ and do not have a spectral type assigned by hand (stars not in Sample 1). In each simulation, we plant a star next to each of these sources between 4 to 6 pixels from the untyped star (red circles). These locations are also chosen so that they will not fall on another source known from imaging or another simulated source. By planting stars next to the untyped sources, we can sample the environmental factors that contribute to the incompleteness, such as halo noise from being near bright sources.}
\label{fig:star_planting_example}
\end{figure}

These complexities result in variations in the sensitivity to early-type and late-type spectra, depending on the location of a star. We determine the relative probability that each untyped source is early-type or late-type by running a series of Monte Carlo simulations. For each untyped star with brightness K$^\prime$ $< 16.0$ mag, we simulate and plant 100 late-type and 100 early-type stars nearby. Each simulated star is planted in the following fashion.
\begin{enumerate}
\item We randomly choose a template spectrum to plant from the catalog of spectra that have already been spectral-typed as early or late-type. The template is required to have SNR $> 35$ to be chosen. The planted young stars are also required to have a measured Br $\gamma$ equivalent width and excludes all WR stars. These criteria are chosen to exclude the spectra from the more massive young stars from the simulations (which may have no Br $\gamma$ absorption). We do not expect the untyped population to include these types of stars as the majority of the untyped stars are much fainter. There are 83 late-type and 41 early-type spectra satisfying these requirements. Figure \ref{fig:star_planting_ew} shows the equivalent widths of the template spectra used for the simulations. 
\item The spectrum is scaled to the flux corresponding to the magnitude of the untyped sources (the conversion between magnitude and flux is empirically calibrated with a power-law fit to the flux determined by \textit{Starfinder} on the OSIRIS cube and the magnitudes determined from NIRC2 imaging). 
\item Photon noise is added to the spectrum.
\item We then plant the star next to the untyped source at a distance randomly chosen between 4 and 6 pixels in radius from the untyped source. The location of the simulated source is also chosen to avoid falling on sources detected in deep imaging. Placing the simulated source close to the untyped source helps to sample the same background location and the same halo noise. See Figure \ref{fig:star_planting_example} for an example of the locations of simulated sources with respect to the untyped sources.
\end{enumerate}

\section{UNBINNED FITTING OF RADIAL SURFACE DENSITY PROFILES USING BAYESIAN ANALYSIS}
\label{append:bayesian}
\begin{figure}[!th]
\center
\includegraphics[angle=90,width=6in]{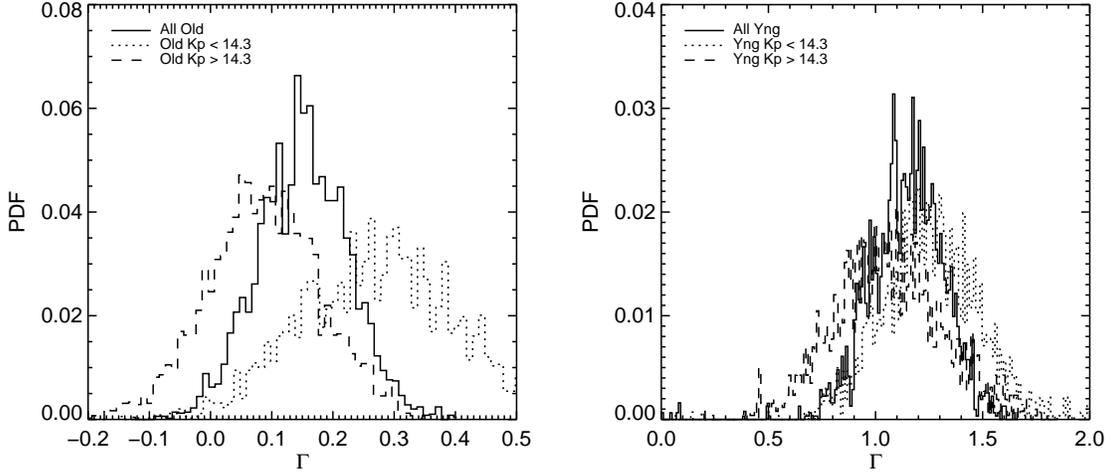}
\caption{The posterior PDFs for the surface-density power-law slope of the late-type  (left) and early-type (right) stars. We also determine the slopes of the sub-population of bright ($K^\prime < 14.3$, dotted) and faint ($K^\prime > 14.3$, dashed) sources. The different sub-populations appear consistent with each other. The most probable values for $\Gamma$ for the different populations are tabulated in Table \ref{tab:gamma}. }
\label{fig:gamma_pdf}
\end{figure}

In order to avoid binning the stellar surface-density profile by radius, we will combine the individual likelihood of each source and compute the posterior distribution for the power-law slope, $\Gamma$ using Bayes' Theorem:
\begin{equation}
P(\Gamma|D) = \frac{P(D|\Gamma)P(\Gamma)}{P(D)}
\end{equation}
where D are the data points, $P(D|\Gamma)$ is the likelihood, $P(\Gamma)$ is the prior distribution of $\Gamma$, and $P(D)$ is the evidence. We will assume a flat prior for $\Gamma$. The surface-density profile is used as the likelihood:
\begin{equation}
\Sigma(x,y,\Gamma) \propto (\sqrt{x^2 + y^2})^{-\Gamma}
\end{equation}
where $x, y$ are the RA and DEC projected positional offsets from Sgr A*. We incorporate the individual positional measurements and their errors by convolving the density profile by a normalized Gaussian ($G(x,y,\sigma{x},\sigma{y})$) centered at the measured position with the error as the $\sigma$. We also normalize the likelihood by integrating over the area of the survey. The likelihood for an individual star is then:
\begin{eqnarray}
\mathcal{L}_i(x_i,y_i|\Gamma,\sigma_{x_i},\sigma_{y_i}) = \frac{\int \Sigma(x,y,\Gamma) G(x_i,y_i|x,y,\sigma_{x_i},\sigma_{y_i}) dx dy }{\int \Sigma(x,y,\Gamma) G(x_i,y_i|x,y,\sigma_{x_i},\sigma_{y_i}) dx dy dx_i dx_i} \\
= \frac{\int \Sigma(x,y,\Gamma) G(x_i,y_i|x,y,\sigma_{x_i},\sigma_{y_i}) dx dy }{\int \Sigma(x,y,\Gamma) dx dy}.
\end{eqnarray}
The membership probability of the star (whether it is young or old) can be easily included by raising the likelihood to the power of the associated probability ($P_E$ or $P_L$); this weighs the likelihood by the star's spectral-type probability. In order to incorporate image completeness, we modify weight by the image completeness at the magnitude of the given star, $I(K^\prime)$, at the field location: $P_{weight} = P_{type}/I(K^\prime)$. The total likelihood is a product of all the individual likelihoods:
\begin{equation}
P(D|\Gamma) = \mathcal{L}_{total} = \prod^N_i \mathcal{L}_i(x_i,y_i|\Gamma,\sigma_{x_i},\sigma_{y_i})^{P_{weight}} 
\end{equation}
The posterior distribution $P(\Gamma|D)$ is then sampled using a Markov-Chain Monte Carlo (MCMC) using the Metropolis-Hastings method. The chains are tested for convergence using the power spectrum method described in \citet{2005MNRAS.356..925D}. The Bayesian analysis leads to a natural way of using all available information in determining the stellar density power-law and is especially useful for determining the properties of subsamples of stars such as splitting the young stars by magnitude. Figure \ref{fig:gamma_pdf} shows the posterior distribution for the completeness-corrected late-type and early-type stars, as well as the subpopulation of faint and bright stars. 

\clearpage

\LongTables
\begin{deluxetable*}{lcccccccccccc}
\tablecolumns{13}
\tablecaption{OSIRIS observations of late-type stars}
\tablewidth{0pc}
\tabletypesize{\scriptsize}  
\tablehead{\colhead{Name} & \colhead{K$^\prime$} & \colhead{$\sigma_{K^\prime}$} & \colhead{K$^\prime_{\Delta A}$} & \colhead{RA offset} & \colhead{DEC offset} & \colhead{R} & \colhead{Epoch} & \colhead{Na}  & \colhead{$\sigma_{Na}$} & $N_{obs}\tablenotemark{b}$ & SNR\tablenotemark{c} & \colhead{Field} \\
\colhead{} & \colhead{} & \colhead{} & \colhead{}& \colhead{(\arcsec)} & \colhead{(\arcsec)} & \colhead{(\arcsec)} & \colhead{}  &
		     \colhead{(\AA)} & \colhead{(\AA)} & \colhead{} & \colhead{} & \colhead{}}
\startdata
\input{old_stars_table}
\enddata
\tablenotetext{a}{Corrected for differential extinction to an $A_{K_s} = 2.7$ from \citet{2010AA...511A..18S}.}
\tablenotetext{b}{Each observation has an integration time of 900 s.}
\tablenotetext{c}{Signal to noise per pixel calculated between 2.212 to 2.218 \micron.}
\tablenotetext{d}{Identified as early-type in \citet{2009ApJ...703.1323D}, but now as late-type. See Appendix \ref{sec:verification} for more details}
\label{table:late}
\end{deluxetable*}

\begin{deluxetable*}{lcccccccccccc}
\tablecolumns{13}
\tablecaption{OSIRIS observations of early-type stars}
\tablewidth{0pc}
\tabletypesize{\scriptsize}  
\tablehead{\colhead{Name} & \colhead{K$^\prime$} & \colhead{$\sigma_{K^\prime}$} & \colhead{K$^\prime_{\Delta A}$\tablenotemark{a}} & \colhead{RA offset} & \colhead{DEC offset} & \colhead{R} & \colhead{Epoch} & \colhead{Br $\gamma$}  & \colhead{$\sigma_{Br \gamma}$} & $N_{obs}\tablenotemark{b}$ & SNR\tablenotemark{c} & \colhead{Field} \\
\colhead{} & \colhead{} & \colhead{} & \colhead{} & \colhead{(\arcsec)} & \colhead{(\arcsec)} & \colhead{(\arcsec)} & \colhead{}  &
		   \colhead{(\AA)} & \colhead{(\AA)} & \colhead{} & \colhead{} & \colhead{}}
\startdata
\input{young_stars_table}
\enddata
\tablenotetext{a}{Corrected for differential extinction to an $A_{K_s} = 2.7$ from \citet{2010AA...511A..18S}}
\tablenotetext{b}{Each observation has an integration time of 900 s.}
\tablenotetext{c}{Signal to noise per pixel calculated between 2.212 to 2.218 \micron.}
\label{table:early}
\end{deluxetable*}

\begin{deluxetable*}{lcccccccc}
\tablecolumns{9}
\tablecaption{Wolf-Rayet stars in the GCOWS field}
\tablewidth{0pc}
\tabletypesize{\scriptsize}  
\tablehead{\colhead{Name} & \colhead{K$^\prime$} & \colhead{$\sigma_{K^\prime}$} & \colhead{K$^\prime_{\Delta A}$\tablenotemark{a}} & \colhead{RA offset} & \colhead{DEC offset} & \colhead{R} & \colhead{Epoch} & \colhead{Field} \\
\colhead{} & \colhead{} & \colhead{} & \colhead{} & \colhead{(\arcsec)} & \colhead{(\arcsec)} & \colhead{(\arcsec)} & \colhead{}  &
		     \colhead{} }
\startdata
\input{wr_stars_table}
\enddata
\tablenotetext{a}{Corrected for differential extinction to an $A_{K_s} = 2.7$ from \citet{2010AA...511A..18S}}
\label{table:wr}
\end{deluxetable*}

\clearpage
\LongTables
\begin{deluxetable*}{lcccccccccccccc}
\tablecolumns{15}
\tablecaption{Completeness Simulation Results}
\tablewidth{0pc}
\tabletypesize{\tiny}  
\tablehead{\colhead{Name} & \colhead{K$^\prime$} & \colhead{$\sigma_{K^\prime}$} & \colhead{K$^\prime_{\Delta A}$} & \colhead{$\Delta$ RA} & \colhead{$\Delta$ Dec} &
           \colhead{$\ln $(BF)\tablenotemark{a}} & \colhead{SNR} &
		   \colhead{Na EW} & \colhead{Na EW Err} &  \colhead{Br $\gamma$ EW}  & \colhead{Br $\gamma$ EW Err} & \colhead{P(Old)\tablenotemark{b}} & \colhead{P(Yng)\tablenotemark{b}} &  \colhead{Field} \\ 
           \colhead{} & \colhead{} & \colhead{} & \colhead{} & \colhead{(\arcsec)} & \colhead{(\arcsec)}  & 
           \colhead{} & \colhead{}  &
           \colhead{(\AA)} & \colhead{(\AA)} & \colhead{(\AA)} & \colhead{(\AA)} & \colhead{} & \colhead{} & \colhead{}}
\startdata
\input{unknownSimTable}
\enddata
\tablenotetext{a}{Blank entries are for stars for which we are unable to measure a Bayes factor, so the probabilities will be the prior. }
\tablenotetext{b}{Probability used in the analysis for this paper. This probability is calculated from Equations \ref{eqn:prob_early} and \ref{eqn:prob_late}, which includes our relative sensitivity to the two types of stars as well as a prior on the relative radial surface-density profiles.}
\label{tab:sim_results}
\end{deluxetable*}
\clearpage
\end{document}

%% file: gcows_field_completeness.tex
9.5-10.0 &	\nodata &	\nodata &	\nodata &	\nodata &	\nodata &	\nodata &	\nodata &	\nodata &	\nodata &	\nodata \\
10.0-10.5 &	\nodata &	\nodata &	\nodata &	1.00 &	\nodata &	\nodata &	\nodata &	\nodata &	\nodata &	1.00 \\
10.5-11.0 &	1.00 &	\nodata &	\nodata &	1.00 &	\nodata &	\nodata &	1.00 &	\nodata &	1.00 &	\nodata \\
11.0-11.5 &	1.00 &	\nodata &	\nodata &	\nodata &	\nodata &	\nodata &	\nodata &	\nodata &	\nodata &	1.00 \\
11.5-12.0 &	\nodata &	\nodata &	1.00 &	\nodata &	\nodata &	\nodata &	1.00 &	\nodata &	\nodata &	\nodata \\
12.0-12.5 &	1.00 &	1.00 &	1.00 &	\nodata &	\nodata &	\nodata &	\nodata &	\nodata &	\nodata &	0.00 \\
12.5-13.0 &	1.00 &	\nodata &	1.00 &	1.00 &	\nodata &	1.00 &	\nodata &	1.00 &	\nodata &	1.00 \\
13.0-13.5 &	1.00 &	1.00 &	1.00 &	\nodata &	1.00 &	1.00 &	1.00 &	1.00 &	1.00 &	0.50 \\
13.5-14.0 &	1.00 &	1.00 &	1.00 &	0.50 &	1.00 &	1.00 &	1.00 &	1.00 &	1.00 &	1.00 \\
14.0-14.5 &	0.00 &	1.00 &	1.00 &	1.00 &	1.00 &	1.00 &	0.67 &	1.00 &	1.00 &	0.60 \\
14.5-15.0 &	0.25 &	1.00 &	1.00 &	1.00 &	1.00 &	1.00 &	1.00 &	1.00 &	0.67 &	0.67 \\
15.0-15.5 &	0.08 &	0.62 &	0.92 &	0.62 &	1.00 &	0.40 &	0.67 &	0.62 &	0.62 &	0.00 \\
15.5-16.0 &	0.00 &	0.00 &	1.00 &	1.00 &	0.50 &	0.33 &	\nodata &	1.00 &	0.67 &	0.00 \\

%% file: verification_sources.tex
S2-317 &	15.52 &	-2.73 &	Late &	Late &	S \\
S2-55 &	15.21 &	-0.68 &	Unknown &	Late &	SE \\
S2-61 &	15.36 &	1.00 &	Unknown &	Late &	E \\
S2-64 &	15.57 &	-1.13 &	Unknown &	Late & E \\
S2-77 &	13.38 &	0.11 &	Late &	Late &	S \\
S3-7 &	13.56 &	-3.50 &	Late &	Late &	SE \\
S3-96 &	14.31 &	-1.96 &	Unknown &	Early &	SW \\
S4-46 &	14.72 &	1.84 &	Unknown &	Late &	E2-2 \\
S5-127 &	15.62 &	-0.41 &	Unknown &	Late &	E2-3 \\
S5-211 &	13.21 &	-1.00 &	Unknown &	Late &	E2-3 \\
S5-237 &	13.21 &	-2.92 &	Unknown &	Early &	E2-1 \\
S6-77 &	14.11 &	-3.13 &	Late &	Late &	E2-3 \\

%% file: old_stars_table.tex
S0-6 	 & 14.10 	 & 0.14 	 & 14.32 	 & 0.03 	 & -0.36 	 & 0.36	 & 2008.30 	 & 5.32 	 & 0.12 	 & 12 	 & 60 	 & C \\ 
S0-18 	 & 15.12 	 & 0.33 	 & 15.32 	 & -0.12 	 & -0.43 	 & 0.45	 & 2009.03 	 & 3.73 	 & 0.59 	 & 11 	 & 28 	 & C \\ 
S0-12 	 & 14.34 	 & 0.03 	 & 14.50 	 & -0.55 	 & 0.41 	 & 0.69	 & 2007.35 	 & 4.64 	 & 0.20 	 & 7 	 & 47 	 & C \\ 
S0-13 	 & 13.35 	 & 0.12 	 & 13.52 	 & 0.55 	 & -0.41 	 & 0.69	 & 2008.34 	 & 4.94 	 & 0.15 	 & 5 	 & 46 	 & C \\ 
S1-5 	 & 12.66 	 & 0.17 	 & 12.76 	 & 0.33 	 & -0.89 	 & 0.95	 & 2008.08 	 & 5.36 	 & 0.05 	 & 3 	 & 58 	 & C \\ 
S1-10 	 & 14.67 	 & 0.04 	 & 14.81 	 & -1.11 	 & -0.02 	 & 1.11	 & 2008.21 	 & 4.29 	 & 0.35 	 & 6 	 & 22 	 & C \\ 
S1-6 	 & 15.35 	 & 0.02 	 & 15.33 	 & -0.96 	 & 0.74 	 & 1.21	 & 2007.87 	 & 5.06 	 & 1.35 	 & 4 	 & 23 	 & C \\ 
S1-34\tablenotemark{d} 	 & 13.12 	 & 0.14 	 & 13.20 	 & 0.87 	 & -0.99 	 & 1.32	 & 2008.31 	 & 0.77 	 & 0.36 	 & 3 	 & 51 	 & C \\

%% file: young_stars_table.tex
S0-2 	 & 14.13 	 & 0.07 	 & 14.39 	 & -0.01 	 & 0.17 	 & 0.17	 & 2007.89 	 & 3.79 	 & 0.29 	 & 24 	 & 74 	 & C \\ 
S0-1 	 & 14.74 	 & 0.07 	 & 14.96 	 & 0.04 	 & -0.26 	 & 0.26	 & 2006.30 	 & 1.13 	 & 0.51 	 & 7 	 & 31 	 & C \\ 
S0-3 	 & 14.54 	 & 0.03 	 & 14.82 	 & 0.34 	 & 0.12 	 & 0.36	 & 2008.39 	 & 2.37 	 & 0.48 	 & 7 	 & 26 	 & C \\ 
S0-5 	 & 15.06 	 & 0.12 	 & 15.27 	 & 0.18 	 & -0.37 	 & 0.41	 & 2007.99 	 & 7.80 	 & 1.10 	 & 9 	 & 15 	 & C \\ 
S0-11 	 & 15.13 	 & 0.04 	 & 15.41 	 & 0.50 	 & -0.06 	 & 0.51	 & 2008.08 	 & 10.20 	 & 7.60 	 & 9 	 & 21 	 & C \\ 
S0-4 	 & 14.31 	 & 0.14 	 & 14.51 	 & 0.42 	 & -0.31 	 & 0.52	 & 2007.62 	 & 3.10 	 & 3.00 	 & 9 	 & 37 	 & C \\ 
S0-9 	 & 14.25 	 & 0.08 	 & 14.42 	 & 0.21 	 & -0.59 	 & 0.62	 & 2008.03 	 & 5.40 	 & 2.00 	 & 7 	 & 38 	 & C \\ 
S0-31 	 & 15.09 	 & 0.10 	 & 15.31 	 & 0.55 	 & 0.45 	 & 0.71	 & 2007.97 	 & 6.01 	 & 1.54 	 & 6 	 & 22 	 & C \\

%% file: wr_stars_table.tex
IRS16C 	 & 9.94 	 & 0.02 	 & 10.10 	 & 1.07 	 & 0.54 	 & 1.19	 & 2007.78 	 & C \\ 
IRS16NW 	 & 10.13 	 & 0.03 	 & 10.14 	 & 0.07 	 & 1.22 	 & 1.22	 & 2007.81 	 & N \\ 
IRS16SW 	 & 10.08 	 & 0.08 	 & 10.15 	 & 1.10 	 & -0.95 	 & 1.45	 & 2007.87 	 & C \\ 
IRS16SW-E 	 & 11.13 	 & 0.01 	 & 11.09 	 & 1.88 	 & -1.12 	 & 2.19	 & 2007.99 	 & E \\ 
S2-16 	 & 11.99 	 & 0.07 	 & 11.69 	 & -1.05 	 & 2.06 	 & 2.31	 & 2007.80 	 & NW \\ 
IRS16NE 	 & 9.14 	 & 0.05 	 & 9.27 	 & 2.89 	 & 0.98 	 & 3.05	 & 2007.84 	 & NE \\ 
S3-5 	 & 12.00 	 & 0.06 	 & 12.01 	 & 2.95 	 & -1.16 	 & 3.17	 & 2008.18 	 & E \\ 
IRS33E 	 & 10.20 	 & 0.08 	 & 10.14 	 & 0.69 	 & -3.13 	 & 3.21	 & 2008.12 	 & S2-1 \\ 
IRS13E4 	 & 11.73 	 & 0.06 	 & 11.65 	 & -3.23 	 & -1.41 	 & 3.52	 & 2008.01 	 & SW \\ 
IRS13E2 	 & 10.69 	 & 0.04 	 & 10.62 	 & -3.19 	 & -1.73 	 & 3.63	 & 2007.84 	 & SW \\ 
S6-82 	 & 13.51 	 & 0.13 	 & 13.48 	 & 6.72 	 & -0.47 	 & 6.73	 & 2008.08 	 & E2-2 \\ 
S9-1 	 & 12.65 	 & 0.01 	 & 12.87 	 & 9.45 	 & 0.28 	 & 9.45	 & 2008.47 	 & E3-1 \\ 

%% file: unknownSimTable.tex
S0-35  &	 15.30 &	 0.12 &	 15.48 &	 0.04 &	 0.90 &	 5.86 &	 17 &	 5.44 &	 1.44 &	 2.86 &	 2.43 &	 1.00 &	 0.00 &	 C \\
S1-27  &	 15.35 &	 0.06 &	 15.43 &	 -1.03 &	 0.20 &	 -3.31 &	 30 &	 -0.13 &	 0.74 &	 2.36 &	 0.45 &	 0.31 &	 0.69 &	 C \\
S1-29  &	 15.28 &	 0.01 &	 15.45 &	 1.08 &	 0.16 &	 4.20 &	 8 &	 6.35 &	 1.85 &	 4.11 &	 1.21 &	 1.00 &	 0.00 &	 C \\
S1-32  &	 15.19 &	 0.12 &	 15.30 &	 -0.98 &	 -0.65 &	 9.66 &	 20 &	 5.30 &	 1.10 &	 0.83 &	 0.36 &	 1.00 &	 0.00 &	 C \\
S1-50  &	 15.36 &	 0.06 &	 15.40 &	 1.49 &	 0.66 &	 3.09 &	 11 &	 4.41 &	 2.01 &	 1.10 &	 0.58 &	 1.00 &	 0.00 &	 C \\
S1-51  &	 14.97 &	 0.07 &	 15.07 &	 -1.66 &	 -0.17 &	 \nodata &	 0 &	 \nodata &	 \nodata &	 \nodata &	 \nodata &	 0.50 &	 0.50 &	 C \\
S1-53  &	 15.20 &	 0.03 &	 15.28 &	 1.67 &	 -0.11 &	 2.76 &	 12 &	 4.14 &	 1.98 &	 1.72 &	 0.55 &	 1.00 &	 0.00 &	 E \\
S1-45  &	 15.21 &	 0.03 &	 15.14 &	 -1.27 &	 1.09 &	 1.75 &	 17 &	 2.46 &	 1.18 &	 2.57 &	 0.65 &	 0.96 &	 0.04 &	 C \\

%% file: ms.bbl
\begin{thebibliography}{57}
\expandafter\ifx\csname natexlab\endcsname\relax\def\natexlab#1{#1}\fi

\bibitem[{{Bahcall} \& {Wolf}(1977)}]{1977ApJ...216..883B}
{Bahcall}, J.~N., \& {Wolf}, R.~A. 1977, \apj, 216, 883

\bibitem[{{Bartko} {et~al.}(2009){Bartko}, {Martins}, {Fritz}, {Genzel},
  {Levin}, {Perets}, {Paumard}, {Nayakshin}, {Gerhard}, {Alexander},
  {Dodds-Eden}, {Eisenhauer}, {Gillessen}, {Mascetti}, {Ott}, {Perrin},
  {Pfuhl}, {Reid}, {Rouan}, {Sternberg}, \& {Trippe}}]{2009ApJ...697.1741B}
{Bartko}, H., {Martins}, F., {Fritz}, T.~K., {et~al.} 2009, \apj, 697, 1741

\bibitem[{{Bartko} {et~al.}(2010){Bartko}, {Martins}, {Trippe}, {Fritz},
  {Genzel}, {Ott}, {Eisenhauer}, {Gillessen}, {Paumard}, {Alexander},
  {Dodds-Eden}, {Gerhard}, {Levin}, {Mascetti}, {Nayakshin}, {Perets},
  {Perrin}, {Pfuhl}, {Reid}, {Rouan}, {Zilka}, \&
  {Sternberg}}]{2010ApJ...708..834B}
{Bartko}, H., {Martins}, F., {Trippe}, S., {et~al.} 2010, \apj, 708, 834

\bibitem[{{Bastian} {et~al.}(2010){Bastian}, {Covey}, \&
  {Meyer}}]{2010ARAA..48..339B}
{Bastian}, N., {Covey}, K.~R., \& {Meyer}, M.~R. 2010, \araa, 48, 339

\bibitem[{{Becklin} \& {Neugebauer}(1968)}]{1968ApJ...151..145B}
{Becklin}, E.~E., \& {Neugebauer}, G. 1968, \apj, 151, 145

\bibitem[{{Blum} {et~al.}(2003){Blum}, {Ram{\'{\i}}rez}, {Sellgren}, \&
  {Olsen}}]{2003ApJ...597..323B}
{Blum}, R.~D., {Ram{\'{\i}}rez}, S.~V., {Sellgren}, K., \& {Olsen}, K. 2003,
  \apj, 597, 323

\bibitem[{{Blum} {et~al.}(1996){Blum}, {Sellgren}, \&
  {Depoy}}]{1996ApJ...470..864B}
{Blum}, R.~D., {Sellgren}, K., \& {Depoy}, D.~L. 1996, \apj, 470, 864

\bibitem[{{Buchholz} {et~al.}(2009){Buchholz}, {Sch{\"o}del}, \&
  {Eckart}}]{2009AA...499..483B}
{Buchholz}, R.~M., {Sch{\"o}del}, R., \& {Eckart}, A. 2009, \aap, 499, 483

\bibitem[{{Depoy} {et~al.}(1993){Depoy}, {Terndrup}, {Frogel}, {Atwood}, \&
  {Blum}}]{1993AJ....105.2121D}
{Depoy}, D.~L., {Terndrup}, D.~M., {Frogel}, J.~A., {Atwood}, B., \& {Blum}, R.
  1993, \aj, 105, 2121

\bibitem[{{Diolaiti} {et~al.}(2000){Diolaiti}, {Bendinelli}, {Bonaccini},
  {Close}, {Currie}, \& {Parmeggiani}}]{2000AAS..147..335D}
{Diolaiti}, E., {Bendinelli}, O., {Bonaccini}, D., {et~al.} 2000, \aaps, 147,
  335

\bibitem[{{Do}(2010)}]{2010PhDTtdo}
{Do}, T. 2010, PhD thesis, University of California, Los Angeles

\bibitem[{{Do} {et~al.}(2009{\natexlab{a}}){Do}, {Ghez}, {Morris}, {Lu},
  {Matthews}, {Yelda}, \& {Larkin}}]{2009ApJ...703.1323D}
{Do}, T., {Ghez}, A.~M., {Morris}, M.~R., {et~al.} 2009{\natexlab{a}}, \apj,
  703, 1323

\bibitem[{{Do} {et~al.}(2009{\natexlab{b}}){Do}, {Ghez}, {Morris}, {Yelda},
  {Meyer}, {Lu}, {Hornstein}, \& {Matthews}}]{2009ApJ...691.1021D}
---. 2009{\natexlab{b}}, \apj, 691, 1021

\bibitem[{{Do} {et~al.}(2012){Do}, {Ghez}, {Lu}, {Morris}, {Yelda}, {Martinez},
  {Peter}, {Wright}, {Bullock}, {Kaplinghat}, \&
  {Matthews}}]{2012JPhCS.372a2016D}
{Do}, T., {Ghez}, A., {Lu}, J.~R., {et~al.} 2012, Journal of Physics Conference
  Series, 372, 012016

\bibitem[{{Ducati} {et~al.}(2001){Ducati}, {Bevilacqua}, {Rembold}, \&
  {Ribeiro}}]{2001ApJ...558..309D}
{Ducati}, J.~R., {Bevilacqua}, C.~M., {Rembold}, S.~B., \& {Ribeiro}, D. 2001,
  \apj, 558, 309

\bibitem[{{Dunkley} {et~al.}(2005){Dunkley}, {Bucher}, {Ferreira}, {Moodley},
  \& {Skordis}}]{2005MNRAS.356..925D}
{Dunkley}, J., {Bucher}, M., {Ferreira}, P.~G., {Moodley}, K., \& {Skordis}, C.
  2005, \mnras, 356, 925

\bibitem[{{Eisenhauer} {et~al.}(2005){Eisenhauer}, {Genzel}, {Alexander},
  {Abuter}, {Paumard}, {Ott}, {Gilbert}, {Gillessen}, {Horrobin}, {Trippe},
  {Bonnet}, {Dumas}, {Hubin}, {Kaufer}, {Kissler-Patig}, {Monnet},
  {Str{\"o}bele}, {Szeifert}, {Eckart}, {Sch{\"o}del}, \&
  {Zucker}}]{2005ApJ...628..246E}
{Eisenhauer}, F., {Genzel}, R., {Alexander}, T., {et~al.} 2005, \apj, 628, 246

\bibitem[{{F{\"o}rster Schreiber}(2000)}]{2000AJ....120.2089F}
{F{\"o}rster Schreiber}, N.~M. 2000, \aj, 120, 2089

\bibitem[{{Frogel} \& {Whitford}(1987)}]{1987ApJ...320..199F}
{Frogel}, J.~A., \& {Whitford}, A.~E. 1987, \apj, 320, 199

\bibitem[{{Genzel} {et~al.}(2010){Genzel}, {Eisenhauer}, \&
  {Gillessen}}]{2010RvMP...82.3121G}
{Genzel}, R., {Eisenhauer}, F., \& {Gillessen}, S. 2010, Reviews of Modern
  Physics, 82, 3121

\bibitem[{{Genzel} {et~al.}(2003){Genzel}, {Sch{\"o}del}, {Ott}, {Eisenhauer},
  {Hofmann}, {Lehnert}, {Eckart}, {Alexander}, {Sternberg}, {Lenzen},
  {Cl{\'e}net}, {Lacombe}, {Rouan}, {Renzini}, \&
  {Tacconi-Garman}}]{2003ApJ...594..812G}
{Genzel}, R., {Sch{\"o}del}, R., {Ott}, T., {et~al.} 2003, \apj, 594, 812

\bibitem[{{Ghez} {et~al.}(2005){Ghez}, {Salim}, {Hornstein}, {Tanner}, {Lu},
  {Morris}, {Becklin}, \& {Duch{\^e}ne}}]{2005ApJ...620..744G}
{Ghez}, A.~M., {Salim}, S., {Hornstein}, S.~D., {et~al.} 2005, \apj, 620, 744

\bibitem[{{Ghez} {et~al.}(2008){Ghez}, {Salim}, {Weinberg}, {Lu}, {Do}, {Dunn},
  {Matthews}, {Morris}, {Yelda}, {Becklin}, {Kremenek}, {Milosavljevic}, \&
  {Naiman}}]{2008ApJ...689.1044G}
{Ghez}, A.~M., {Salim}, S., {Weinberg}, N.~N., {et~al.} 2008, \apj, 689, 1044

\bibitem[{{Gillessen} {et~al.}(2009){Gillessen}, {Eisenhauer}, {Trippe},
  {Alexander}, {Genzel}, {Martins}, \& {Ott}}]{2009ApJ...692.1075G}
{Gillessen}, S., {Eisenhauer}, F., {Trippe}, S., {et~al.} 2009, \apj, 692, 1075

\bibitem[{{Haller} {et~al.}(1996){Haller}, {Rieke}, {Rieke}, {Tamblyn},
  {Close}, \& {Melia}}]{1996ApJ...456..194H}
{Haller}, J.~W., {Rieke}, M.~J., {Rieke}, G.~H., {et~al.} 1996, \apj, 456, 194

\bibitem[{{Hanson} {et~al.}(1996){Hanson}, {Conti}, \&
  {Rieke}}]{1996ApJS..107..281H}
{Hanson}, M.~M., {Conti}, P.~S., \& {Rieke}, M.~J. 1996, \apjs, 107, 281

\bibitem[{{Krabbe} {et~al.}(1991){Krabbe}, {Genzel}, {Drapatz}, \&
  {Rotaciuc}}]{1991ApJ...382L..19K}
{Krabbe}, A., {Genzel}, R., {Drapatz}, S., \& {Rotaciuc}, V. 1991, \apjl, 382,
  L19

\bibitem[{{Larkin} {et~al.}(2006){Larkin}, {Barczys}, {Krabbe}, {Adkins},
  {Aliado}, {Amico}, {Brims}, {Campbell}, {Canfield}, {Gasaway}, {Honey},
  {Iserlohe}, {Johnson}, {Kress}, {Lafreniere}, {Magnone}, {Magnone},
  {McElwain}, {Moon}, {Quirrenbach}, {Skulason}, {Song}, {Spencer}, {Weiss}, \&
  {Wright}}]{2006NewAR..50..362L}
{Larkin}, J., {Barczys}, M., {Krabbe}, A., {et~al.} 2006, New Astronomy Review,
  50, 362

\bibitem[{{Levin} \& {Beloborodov}(2003)}]{2003ApJ...590L..33L}
{Levin}, Y., \& {Beloborodov}, A.~M. 2003, \apjl, 590, L33

\bibitem[{{Lu}(2008)}]{luThesis}
{Lu}, J.~R. 2008, PhD thesis, UCLA

\bibitem[{{Lu} {et~al.}(2012){Lu}, {Do}, {Ghez}, {Morris}, {Yelda}, \&
  {Matthews}}]{lu2012}
{Lu}, J.~R., {Do}, T., {Ghez}, A.~M., {et~al.} 2012, in preparation

\bibitem[{{Lu} {et~al.}(2009){Lu}, {Ghez}, {Hornstein}, {Morris}, {Becklin}, \&
  {Matthews}}]{2009ApJ...690.1463L}
{Lu}, J.~R., {Ghez}, A.~M., {Hornstein}, S.~D., {et~al.} 2009, \apj, 690, 1463

\bibitem[{{Madigan} {et~al.}(2009){Madigan}, {Levin}, \&
  {Hopman}}]{2009ApJ...697L..44M}
{Madigan}, A., {Levin}, Y., \& {Hopman}, C. 2009, \apjl, 697, L44

\bibitem[{{Madigan} {et~al.}(2011){Madigan}, {Hopman}, \&
  {Levin}}]{2011ApJ...738...99M}
{Madigan}, A.-M., {Hopman}, C., \& {Levin}, Y. 2011, \apj, 738, 99

\bibitem[{{Maness} {et~al.}(2007){Maness}, {Martins}, {Trippe}, {Genzel},
  {Graham}, {Sheehy}, {Salaris}, {Gillessen}, {Alexander}, {Paumard}, {Ott},
  {Abuter}, \& {Eisenhauer}}]{2007ApJ...669.1024M}
{Maness}, H., {Martins}, F., {Trippe}, S., {et~al.} 2007, \apj, 669, 1024

\bibitem[{{Martins} {et~al.}(2007){Martins}, {Genzel}, {Hillier}, {Eisenhauer},
  {Paumard}, {Gillessen}, {Ott}, \& {Trippe}}]{2007AA...468..233M}
{Martins}, F., {Genzel}, R., {Hillier}, D.~J., {et~al.} 2007, \aap, 468, 233

\bibitem[{{Martins} \& {Plez}(2006)}]{2006AA...457..637M}
{Martins}, F., \& {Plez}, B. 2006, \aap, 457, 637

\bibitem[{{McKee} \& {Ostriker}(2007)}]{2007ARAA..45..565M}
{McKee}, C.~F., \& {Ostriker}, E.~C. 2007, \araa, 45, 565

\bibitem[{{Merritt}(2010)}]{2010ApJ...718..739M}
{Merritt}, D. 2010, \apj, 718, 739

\bibitem[{{Murphy}(2011)}]{2011ASPC..439..189M}
{Murphy}, B.~W. 2011, in Astronomical Society of the Pacific Conference Series,
  Vol. 439, The Galactic Center: a Window to the Nuclear Environment of Disk
  Galaxies, ed. {M.~R.~Morris, Q.~D.~Wang, \& F.~Yuan}, 189--+

\bibitem[{{Nishiyama} {et~al.}(2009){Nishiyama}, {Tamura}, {Hatano}, {Kato},
  {Tanab{\'e}}, {Sugitani}, \& {Nagata}}]{2009ApJ...696.1407N}
{Nishiyama}, S., {Tamura}, M., {Hatano}, H., {et~al.} 2009, \apj, 696, 1407

\bibitem[{{Paumard} {et~al.}(2004){Paumard}, {Maillard}, \&
  {Morris}}]{2004AA...426...81P}
{Paumard}, T., {Maillard}, J.-P., \& {Morris}, M. 2004, \aap, 426, 81

\bibitem[{{Paumard} {et~al.}(2006){Paumard}, {Genzel}, {Martins}, {Nayakshin},
  {Beloborodov}, {Levin}, {Trippe}, {Eisenhauer}, {Ott}, {Gillessen}, {Abuter},
  {Cuadra}, {Alexander}, \& {Sternberg}}]{2006ApJ...643.1011P}
{Paumard}, T., {Genzel}, R., {Martins}, F., {et~al.} 2006, \apj, 643, 1011

\bibitem[{{Pfuhl} {et~al.}(2011){Pfuhl}, {Fritz}, {Zilka}, {Maness},
  {Eisenhauer}, {Genzel}, {Gillessen}, {Ott}, {Dodds-Eden}, \&
  {Sternberg}}]{2011ApJ...741..108P}
{Pfuhl}, O., {Fritz}, T.~K., {Zilka}, M., {et~al.} 2011, \apj, 741, 108

\bibitem[{{Preto} \& {Amaro-Seoane}(2010)}]{2010ApJ...708L..42P}
{Preto}, M., \& {Amaro-Seoane}, P. 2010, \apjl, 708, L42

\bibitem[{{Rayner} {et~al.}(2009){Rayner}, {Cushing}, \&
  {Vacca}}]{2009ApJS..185..289R}
{Rayner}, J.~T., {Cushing}, M.~C., \& {Vacca}, W.~D. 2009, \apjs, 185, 289

\bibitem[{{Sch{\"o}del}(2011)}]{2011ASPC..439..222S}
{Sch{\"o}del}, R. 2011, in Astronomical Society of the Pacific Conference
  Series, Vol. 439, The Galactic Center: a Window to the Nuclear Environment of
  Disk Galaxies, ed. {M.~R.~Morris, Q.~D.~Wang, \& F.~Yuan}, 222--+

\bibitem[{{Sch{\"o}del} {et~al.}(2010){Sch{\"o}del}, {Najarro}, {Muzic}, \&
  {Eckart}}]{2010AA...511A..18S}
{Sch{\"o}del}, R., {Najarro}, F., {Muzic}, K., \& {Eckart}, A. 2010, \aap, 511,
  A18+

\bibitem[{{Sch{\"o}del} {et~al.}(2007){Sch{\"o}del}, {Eckart}, {Alexander},
  {Merritt}, {Genzel}, {Sternberg}, {Meyer}, {Kul}, {Moultaka}, {Ott}, \&
  {Straubmeier}}]{2007AA...469..125S}
{Sch{\"o}del}, R., {Eckart}, A., {Alexander}, T., {et~al.} 2007, \aap, 469, 125

\bibitem[{{Tiede} {et~al.}(1995){Tiede}, {Frogel}, \&
  {Terndrup}}]{1995AJ....110.2788T}
{Tiede}, G.~P., {Frogel}, J.~A., \& {Terndrup}, D.~M. 1995, \aj, 110, 2788

\bibitem[{{Trotta}(2008)}]{2008ConPh..49...71T}
{Trotta}, R. 2008, Contemporary Physics, 49, 71

\bibitem[{{van Dam} {et~al.}(2006){van Dam}, {Bouchez}, {Le Mignant},
  {Johansson}, {Wizinowich}, {Campbell}, {Chin}, {Hartman}, {Lafon}, {Stomski},
  \& {Summers}}]{2006PASP..118..310V}
{van Dam}, M.~A., {Bouchez}, A.~H., {Le Mignant}, D., {et~al.} 2006, \pasp,
  118, 310

\bibitem[{{Wegner}(2007)}]{2007MNRAS.374.1549W}
{Wegner}, W. 2007, \mnras, 374, 1549

\bibitem[{{Williams} \& {Antonopoulou}(1981)}]{1981MNRAS.196..915W}
{Williams}, P.~M., \& {Antonopoulou}, E. 1981, \mnras, 196, 915

\bibitem[{{Wizinowich} {et~al.}(2006){Wizinowich}, {Le Mignant}, {Bouchez},
  {Campbell}, {Chin}, {Contos}, {van Dam}, {Hartman}, {Johansson}, {Lafon},
  {Lewis}, {Stomski}, {Summers}, {Brown}, {Danforth}, {Max}, \&
  {Pennington}}]{2006PASP..118..297W}
{Wizinowich}, P.~L., {Le Mignant}, D., {Bouchez}, A.~H., {et~al.} 2006, \pasp,
  118, 297

\bibitem[{{Yelda}(2012)}]{yeldaThesis}
{Yelda}, S. 2012, PhD thesis, UCLA

\bibitem[{{Yelda} {et~al.}(2010){Yelda}, {Lu}, {Ghez}, {Clarkson}, {Anderson},
  {Do}, \& {Matthews}}]{2010ApJ...725..331Y}
{Yelda}, S., {Lu}, J.~R., {Ghez}, A.~M., {et~al.} 2010, \apj, 725, 331

\end{thebibliography}
